\documentclass[floatfix,aps,prb,twocolumn]{revtex4-2}

\usepackage{comment}
\usepackage[T1]{fontenc}
\usepackage{lipsum}  
\usepackage{graphicx}
\usepackage[english]{babel}
\usepackage{upgreek}
\usepackage{amsmath}
\usepackage{amsfonts}
\usepackage{physics}
\usepackage{array}
\usepackage[colorlinks=true,citecolor=blue,linkcolor=blue,urlcolor=blue]{hyperref}
\usepackage[dvipsnames]{xcolor}
\usepackage[normalem]{ulem}
\usepackage{hhline}
\usepackage{bbold}
\usepackage{float}
\usepackage{tikz}
\usetikzlibrary{positioning}

\makeatletter
\def\bbl@set@language#1{%
	\edef\languagename{%
		\ifnum\escapechar=\expandafter`\string#1\@empty
		\else\string#1\@empty\fi}%
	\@ifundefined{babel@language@alias@\languagename}{}{%
		\edef\languagename{\@nameuse{babel@language@alias@\languagename}}%
	}%
	\select@language{\languagename}%
	\expandafter\ifx\csname date\languagename\endcsname\relax\else
	\if@filesw
	\protected@write\@auxout{}{\string\select@language{\languagename}}%
	\bbl@for\bbl@tempa\BabelContentsFiles{%
		\addtocontents{\bbl@tempa}{\xstring\select@language{\languagename}}}%
	\bbl@usehooks{write}{}%
	\fi
	\fi}
\newcommand{\DeclareLanguageAlias}[2]{%
	\global\@namedef{babel@language@alias@#1}{#2}%
}
\makeatother

\DeclareLanguageAlias{en}{english}

\newcommand\varpm{\mathbin{\vcenter{\hbox{%
				\oalign{\hfil$\scriptstyle+$\hfil\cr
					\noalign{\kern-.3ex}
					$\scriptscriptstyle({-})$\cr}%
}}}}
\newcommand\varmp{\mathbin{\vcenter{\hbox{%
				\oalign{$\scriptstyle({+})$\cr
					\noalign{\kern-.3ex}
					\hfil$\scriptscriptstyle-$\hfil\cr}%
}}}}

\newcommand{\bigzero}{\mbox{\normalfont\Large\bfseries 0}}

\begin{document}


\title{Dispersive cavity-mediated quantum gate between driven dot-donor nuclear spins}




\author{Jonas Mielke}
\author{Guido Burkard}
\affiliation{Department of Physics, University of Konstanz, Konstanz D-78457, Germany}



\begin{abstract}
Nuclear spins show exceptionally long coherence times but the underlying good isolation from their environment is a challenge when it comes to controlling nuclear spin qubits. A particular difficulty, not only for nuclear spin qubits, is the realization of two-qubit gates between distant qubits. Recently, strong coupling between an electron spin and microwave resonator photons  as well as a microwave resonator mediated coupling between two electron spins both in the resonant and the dispersive regime have been reported and, thus, a microwave resonator mediated electron spin two qubit gate seems to be in reach.  Inspired by these findings, we theoretically investigate the interaction of a microwave resonator with a hybrid quantum dot-donor (QDD) system consisting of a gate defined Si QD and a laterally displaced $^{31}$P phosphorous donor atom implanted in the Si host material. We find that driving the QDD system allows to compensate the frequency mismatch between the donor nuclear spin splitting in the MHz regime and typical superconducting  resonator frequencies in the GHz regime, and also enables an effective nuclear spin-photon coupling. While we expect this coupling to be weak, we predict that coupling the nuclear spins of two distant QDD systems dispersively to the microwave resonator allows the implementation of a resonator mediated  nuclear spin two-qubit $\sqrt{i\mathrm{SWAP}}$ gate with a gate fidelity approaching $90\%$.
\end{abstract}


\maketitle

\section{Introduction}

The exceptionally long coherence times  reported for nuclear spins \cite{steger2012,pla2013,saeedi2013,muhonen2014} suggest the large potential of  nuclear spins as qubit implementations for quantum information applications, which is why some early proposals for quantum computers are based on nuclear spins \cite{kane1998}. 

Considering spin-based quantum computing architectures, so far, mostly the fundamental elements of a quantum computer have been realized in experiments. This involves, in particular, the realization of one and two-qubit devices with which high-fidelity one and two-qubit gates were demonstrated. However, recently, also linear and two-dimensional arrays of qubits  were fabricated and the control of the individual qubits was achieved \cite{lawrie2020a,hendrickx2021,philips2022a}. This accomplishment constitutes an important step towards larger setups in the near future. With such systems in reach and aiming at a universal quantum computer with a large number of qubits, the connectivity of the qubits, i.e. the pairs of physically coupled qubits, is of great importance and, ideally, all-to-all qubit connectivity is achieved. Exploiting the exchange coupling between neighboring qubits has proven successful for the implementation of two-qubit gates between hole spin qubits \cite{hendrickx2021} and the realization of high fidelity two-qubit gates between electron spin qubits exceeding the surface code threshold \cite{noiri2022,xue2022,mills2022}. A nuclear spin two-qubit gate with such a high fidelity is obtained between the nuclear spins of neighboring phosphorous donor atoms in a configuration where the spins are hyperfine-coupled to the same electron \cite{madzik2022}. However, the underlying interactions are short-ranged and do not allow for the coupling of distant qubits. Harnessing an intermediate system as a mediator for an interaction is an approach to deal with this obstacle. 

In case of electron spins microwave resonators are promising candidates for such an intermediate system. A single electron confined in a double quantum dot (DQD) acquires a large electric dipole moment \cite{burkard2020}. Applying a magnetic field gradient between the two quantum dots (QD) forming the DQD allows to engineer an artificial spin orbit interaction, that, in turn, enables an effective coupling between the electron spin and microwave resonator photons \cite{benito2017,burkard2020}.  Notably, the strong electron spin-photon coupling regime has been attained \cite{mi2018,samkharadze2018} and, thereby paved the way for experiments demonstrating the microwave resonator photon mediated interaction between spin qubits both in the resonant \cite{borjans2020} and the dispersive \cite{harvey-collard2022} coupling regime. The latter achievement is particularly promising with regard to the theoretical proposal for a dispersive cavity mediated $i\mathrm{SWAP}$ gate \cite{benito2019a}. 
Theory also predicts that strong spin-photon coupling can be achieved with pairs of donors in Si \cite{osika2022}.

Looking at nuclear spins, a hybrid architecture in which a single electron is shared between an interfacial quantum dot and an implanted $^{31}$P donor atom beneath it permits the implementation of an electron spin-nuclear spin flip-flop qubit with the two qubit states corresponding to the anti-parallel states of electron and nuclear spins \cite{tosi2017,savytskyy2023}. The spatial displacement of the electron wave function away from the donor gives rise to an electric dipole, such that the electron spin-nuclear spin flip-flop qubit is expected to couple to a microwave resonator \cite{tosi2017}. With the help of ac-magnetic fields, this system can also be harnessed to couple the nuclear spin to microwave cavity photons \cite{tosi2018}. However, the usage of oscillating magnetic fields is unfavorable in case of many-qubit devices because spatial localization at the nanoscale is not possible and, thus, nuclear spin-photon coupling by all-electrical means is desired. 

With this challenge in mind and inspired by the aforementioned findings for the DQD system and the hybrid vertical architecture, we theoretically investigate a hybrid quantum dot-donor (QDD) architecture composed of a gate defined Si QD and a thereto laterally displaced $^{31}$P donor atom implanted in the Si host material of the quantum well. 
While such a device has successfully been operated in the multi-electron regime \cite{harvey-collard2017}, our analysis is restricted to the single electron scenario. In addition to the setup in the experiment, but  similar to the DQD system mentioned before, the architecture envisioned here includes a magnetic field gradient that causes an effective electron spin-orbit coupling. Then, due to the similarity with the DQD system, an effective electron spin-photon coupling is expected and has been theoretically verified \cite{mielke2021}. The special feature of the considered setup is the hyperfine interaction between the nuclear spin and the electron spin. While this interaction underlies the theoretically suggested method for nuclear spin state readout by probing the microwave resonator transmission, it does not allow for resonant nuclear spin-photon coupling due to the large frequency mismatch between the nuclear spin transition frequency of the order of MHz and the resonator photons in the microwave domain without further modifications of the system \cite{mielke2021}.

 Here, we demonstrate that the frequency mismatch can be compensated by periodically driving the QDD detuning away from the point where the electron is fully hybridized between the QD and the donor.  Our theoretical analysis unveils that the strong nuclear spin-photon coupling regime is out of reach assuming state-of-the-art device parameters, and, therefore, the coherent excitation exchange between the nuclear spin states and the microwave resonator is not possible. Nonetheless, the nuclear spin-photon coupling can be exploited for quantum information applications by dispersively coupling the nuclear spins of two driven QDD systems to a microwave resonator. For such a setup, we find an effective interaction between the two nuclear spins mediated by virtual resonator photons. This interaction can be harnessed to implement either a nuclear spin  $\sqrt{i\mathrm{SWAP}}$ or $i\mathrm{SWAP}$ quantum gate. Our sophisticated numerical simulations accounting for decoherence effects suggest that average gate fidelities approaching 90\% and 80\%, respectively, can be achieved assuming realistic system parameters, state-of-the-art decay and decoherence times reported for charge and spin qubits, and a resonator with a high quality factor.
 
This article is structured as follows. Section~\ref{sec:model} provides a detailed description of the model employed to describe a driven QDD system interacting with a microwave resonator. In Sec.~\ref{sec:nuclear spin-photon coupling} we theoretically derive an effective nuclear spin-photon coupling, investigate its strength, and thereby elaborate on accessible coupling regimes. Then, in Sec.~\ref{sec:dispersive_nuclear_spin_gate}, two driven QDD systems dispersively coupled to a common microwave resonator are investigated, an effective photon-mediated coupling between the nuclear spins of the QDD systems is found, and it is demonstrated that the coupling allows the implementation of an $\sqrt{i\mathrm{SWAP}}$ or an $i\mathrm{SWAP}$ quantum gate even in the presence of decoherence effects. Finally, our results are summarized in Sec.~\ref{sec:conclusion}. 

\section{Quantum dot-donor system \label{sec:model}}
\begin{figure}
	\centering
	\includegraphics[width=\columnwidth]{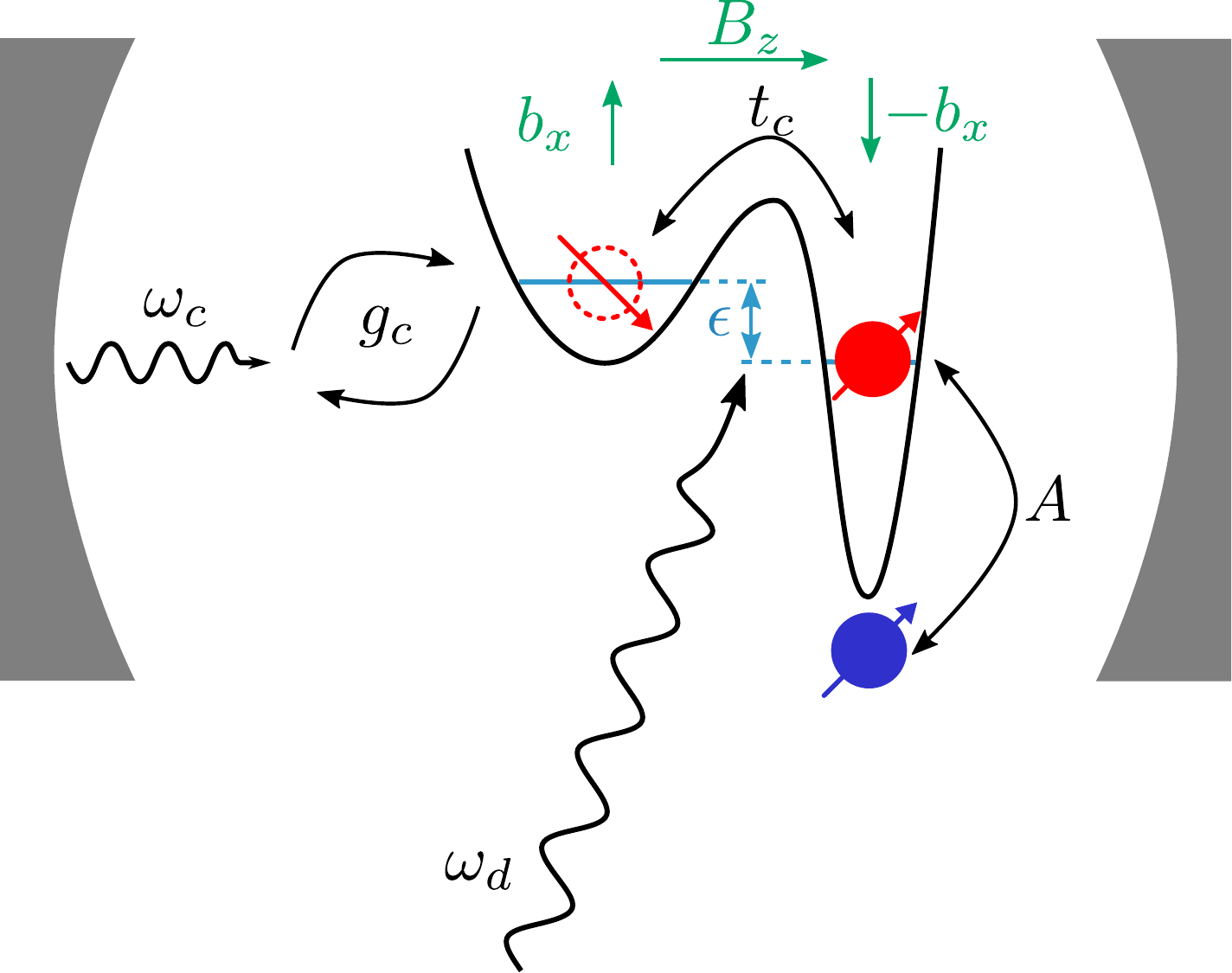}
	\caption{Schematic illustration of the hybrid quantum dot-donor system, populated by a single electron (red), interacting with a single mode $\omega_c$ of a microwave resonator. The relative position of the lowest quantum dot and the lowest donor energy level is characterized by the detuning parameter $\epsilon$. Quantum dot and donor atom are tunnel coupled with coupling strength $t_c$. The spin of the confined electron is subject to a homogeneous magnetic field $B_z$ and it experiences a magnetic field gradient $b_x$ along the quantum dot-donor axis ($z$-axis). The quantum dot-donor system's electric dipole moment couples to the resonator mode with  electric dipole coupling strength $g_c$. Electron spin and donor nuclear spin (blue) interact via the hyperfine interaction if the electron overlaps with the donor atom.}
	\label{fig:quantumdot-donor system}
\end{figure}

We consider a lateral hybrid QDD architecture realized in a Si/SiGe heterostructure with an isotopically purified $^{28}$Si quantum well, see Fig.~\ref{fig:quantumdot-donor system}.
\begin{figure}
    \centering
    \includegraphics[width=\columnwidth]{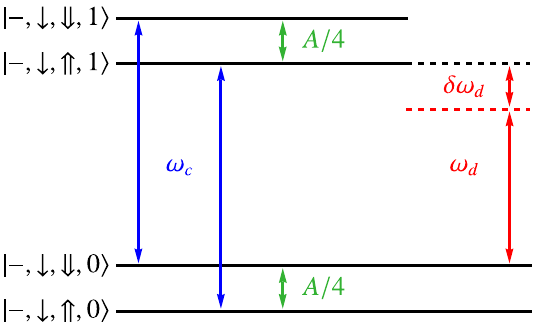}
    \caption{\textcolor{Black}{Schematic energy level diagram of the QDD system coupled to a microwave resonator restricted to the lower charge state $(-)$ and the electron spin ground state $(\downarrow)$ for $0$ and $1$ cavity photons in the lab  (non-rotating) frame. According to \eqref{eq:HQDD} the opposite nuclear spin states are split by the hyperfine interaction $A/4$ (green arrows), while the energy difference between the photon states is set by the resonator frequency $\omega_c$ (blue arrows) with $\omega_c\gg A/4$. The red arrow suggests that a drive with frequency $\omega_d$ can tune the nuclear spin splitting and the photon splitting into resonance $(\delta\omega_d=0)$ or close to resonance $(\delta\omega_d\neq0)$. The successive entries of the states labeling the energy levels represent the electron's orbital state ($-,+$), the electron spin state ($\downarrow,\uparrow$), the nuclear spin state ($\Downarrow,\Uparrow$) and the number of photons in the microwave resonator ($0,1,2,\dotsc$).}}
    \label{fig:leveldiagram}
\end{figure}

 The two constituents of the QDD system are a gate defined QD and an ionized phosphorous donor atom that is  implanted, laterally displaced by 30-40~nm with respect to the QD (here along the $z$-axis), in the Si host material of the quantum well. An electric field applied along  the QDD axis and the gate electrodes generating the QD confinement allow to control the relative position of the lowest QD and donor energy level \cite{harvey-collard2017,mielke2021} that enters our model via the detuning parameter $\epsilon$.  The system can be set up such that there is a sizable tunnel coupling $t_c$ between the QD and the donor site \cite{harvey-collard2017,mielke2021}.  We consider a single electron confined in the QDD potential, that can be mostly confined at the position of the QD ($\epsilon\ll -t_c$), $\ket{L}$, at the position of the donor atom ($\epsilon\gg t_c$), $\ket{R}$, or it is hybridized between the two ($|\epsilon|\lesssim t_c$) due to the tunnel coupling.
 
  The electron spin degeneracy is lifted by a homogeneous magnetic field $B_z$ applied in $z$-direction, while there is also a magnetic field gradient $\partial B_x/\partial z$ generated by a micromagnet with $b_x$ the field strength difference between the QD and the donor site. 
 
  The phosphorous donor atom has a nuclear spin 1/2 in the otherwise nuclear spin free environment of the quantum well. This nuclear spin causes a unique feature that distinguishes the QDD system from DQDs: if the electron wave function overlaps with the donor atom, electron spin and donor nuclear spin interact via the hyperfine interaction, whereby the dominant contribution is due to the contact hyperfine interaction with interaction strength $A$. It is noteworthy that the hyperfine interaction strength $A/2\pi\approx 25\text{ MHz}$  for a phosphorous donor atom implanted in the Si quantum well of a $\mathrm{Si}/\mathrm{Si}_{0.7}\mathrm{Ge}_{0.3}$ heterostructure  is considerably reduced compared to the value known for bulk Si, $A/2\pi=117\text{ MHz}$ \cite{feher1959,steger2011}, due strain effects resulting from the Si- and SiGe- lattice mismatch \cite{usman2015,huebl2006}.  
  
  The described QDD system is modeled by the Hamiltonian
	\begin{align}
	\widetilde{H}_{\mathrm{QDD}}=&\frac{1}{2}\left(\epsilon \widetilde{\tau}_z +2 t_c\widetilde{\tau}_x+B_z \sigma_z+b_x\sigma_x\widetilde{\tau}_z\right)\nonumber\\
	&+\frac{A}{8}\vec{\sigma}\cdot \vec{\nu}\, (1-\widetilde{\tau}_z) ,
	\label{eq:HtildeQDD}
\end{align}
with the three sets of Pauli operators $\{\widetilde{\tau}_i,\,\sigma_i,\,\nu_i\}$. The operators $\widetilde{\tau}_i$ act on the electron position space as  $\widetilde{\tau}_z\ket{L(R)}=\textcolor{Black}{\varpm} \ket{L(R)}$, $\sigma_i$ and $\nu_i$ are the electron spin and nuclear spin Pauli operators, respectively. Here, energy units are chosen such that $\hbar=1$ and the magnetic fields $B_z$ and $b_x$ are given in units of energy. The nuclear spin interaction with the magnetic field is neglected in the above expression because it is typically three orders of magnitude smaller than all other energy scales.

The spatial separation of  the QD and the donor ensures that the confined electron acquires a substantial charge dipole moment, and, thus, couples to the electric field, ${E_{\mathrm{cav}}=E_0\left(a+a^{\dagger}\right)}$, of a microwave resonator \cite{burkard2020,burkard2021}
\begin{align}
	\widetilde{H}_{\mathrm{int}}&=g_c\widetilde{\tau}_z(a+a^{\dagger}), 
	\label{eq:Htildeint}
\end{align}
with $a^{\dagger}$ and $a$ the cavity photon creation and annihilation operators of the relevant cavity mode with frequency $\omega_c$, ${H_{\mathrm{cav}}=\omega_c a^{\dagger}a}$. The charge-photon interaction strength ${g_c=e E_0 d}$ is determined by the cavity vacuum electric field $E_0$ and the QDD distance $d$.

Modulating the voltage applied to the gates defining the QD potential or the electric field applied along the QDD axis periodically with frequency $\omega_d$ results in a drive of the QDD detuning
\begin{align}
	\widetilde{H}_{d}=\frac{\epsilon_d}{2}\cos\left(\omega_d t\right)\widetilde{\tau}_z,
	\label{eq:Htilded}
\end{align}
with drive amplitude $\epsilon_d$.

In line with recent experiments we assume  $2t_c,\,B_z\gg A,\,b_x$  \cite{mi2018} in this work. Therefore, it is more convenient to express the electron position in terms of the eigenstates of the first two terms of \eqref{eq:HtildeQDD}  that resemble bonding $(-)$ and antibonding $(+)$ molecular orbital states,
\begin{align}
&	\ket{\varpm}=\frac{(\sin(\theta)\varpm 1)|L\rangle+\cos(\theta) |R\rangle}{\sqrt{2\varpm2\sin(\theta)}},
\end{align}
with orbital mixing angle $\theta=\arctan(\epsilon /2 t_c)$, and corresponding Pauli operators $\tau_i$ defined by 	$\tau_z\ket{\varpm}=\varpm \ket{\varpm}$. The orbital energies $E_{\varpm}=\varpm\sqrt{\epsilon^2+
	4t_c^2}/2$ are to first order insensitive to charge noise if the system is operated at the charge noise sweet spot characterized by zero QDD detuning, i.e. $\partial E_{\scriptsize{
		\varpm}}/ \partial \epsilon|_{\epsilon=0}=0$. Hence, we consider $\epsilon=0$ in the remainder of this article. For this specific operation point, the electron position operators $\widetilde{\tau}_i$ transform as 
\begin{align}
	&\widetilde{\tau}_x\rightarrow \tau_z,\\
	&\widetilde{\tau}_z \rightarrow -\tau_x.
\end{align}
Additionally assuming that the system is operated in a regime characterized by $A\ll 2t_c\approx B_z\approx \omega_c\approx{\omega_d}$, it is justified to apply the rotating wave approximation (RWA), where \eqref{eq:HtildeQDD}-\eqref{eq:Htilded} simplify to 
	\begin{align}
	H_{\mathrm{QDD}}=&\frac{1}{2}\left[2t_c\tau_z+B_z \sigma_z +\frac{A}{4}\sigma_z\nu_z\right. \nonumber\\
 &\left.-b_x\left(\sigma_+\tau_-+\sigma_-\tau_+\right)\right]\nonumber\\
	&+\frac{A}{4}\left(\sigma_+\nu_{-}\tau_-+\sigma_-\nu_{+}\tau_+\right),
	\label{eq:HQDD}\\
	H_{\mathrm{int}}&=-g_c\left(\tau_+a+\tau_-a^{\dagger}\right),\\
	H_{d}(t)&=-\frac{\epsilon_d}{4} \left(e^{-i\omega_d t}\tau_++e^{i\omega_d t}\tau_-\right).
	\label{eq:Hd}
\end{align}

A further transformation to a rotating reference frame defined by
\begin{align}
	U_{\mathrm{R}}(t)=	U_{\mathrm{R,sys}}(t)U_{\mathrm{R,}a}(t),
	\label{eq:UR}
\end{align}
with
\begin{align}
U_{\mathrm{R,sys}}(t)=&	\exp\left[i \frac{\omega_d}{2}\left( \tau_z+\sigma_z \right)t\right],
\label{eq:URsys}\\
U_{\mathrm{R,}a}(t)=&\exp\left[i \omega_da^{\dagger} at\right],
\label{eq:URa}
\end{align}
allows a time-independent description of the driven system 
\begin{align}
	H^{\mathrm{R}}=&	U_{\mathrm{R}}(t) \left(H_{\mathrm{QDD}}+	H_{\mathrm{int}}+H_{\mathrm{d}}(t)+H_{\mathrm{cav}}\right)U^{\dagger}_{\mathrm{R}}(t)\nonumber\\
	&+i\, \dot{U}_{\mathrm{R}}(t)U^{\dagger}_{\mathrm{R}}(t)\nonumber\\
	=&H^{\mathrm{R}}_{0}+V^{\mathrm{R}},
	\label{eq:HR}
\end{align}
composed of a diagonal part $H^{\mathrm{ R}}_{0}$ and an off-diagonal part $V^{\mathrm{R}}$,
\begin{align}
	H^{\mathrm{ R}}_{0}=&\frac{2 t_c-\omega_d}{2}\tau_z+\frac{B_z-\omega_d}{2} \sigma_z+\frac{A}{8}\sigma_z\nu_{z}\nonumber\\
	&+\left(\omega_c-\omega_d\right) a^{\dagger}a,\\
	V^{\mathrm{R}}=&-\frac{b_x}{2}\left(\sigma_+ \tau_- +\sigma_- \tau_+\right)-g_c\left(\tau_+a+\tau_-a^{\dagger}\right)\nonumber\\
	&+\frac{A}{4}\left(\sigma_+ \nu_- \tau_- +\sigma_- \nu_+\tau_+\right)-\frac{\epsilon_d}{4}\left(\tau_+ +\tau_-\right).
	\label{eq:VR}
\end{align}

\section{Nuclear spin photon coupling \label{sec:nuclear spin-photon coupling}}

A transverse nuclear spin-photon coupling has already been theoretically predicted for a QDD system without the periodic driving of the detuning \cite{mielke2021}. However, the different energy scales of the nuclear spin splitting ($\approx A/4$) and the resonator photons $(\omega_c)$ do not permit coherent excitation exchange\textcolor{Black}{, as illustrated in Fig.~\ref{fig:leveldiagram}}. In this section, we demonstrate that driving the QDD detuning allows to compensate for the frequency mismatch and derive an effective nuclear spin-photon interaction Hamiltonian.  

It is instructive to inspect the energy expectation values of $H^{\mathrm{R}}_0$ with respect to the basis states $\ket{-/+,\downarrow/\uparrow,\Downarrow/\Uparrow,n}$, where the successive entries give the electron's orbital state, the electron spin state, the nuclear spin state and the number of photons in the microwave resonator. In particular, one finds that  $\ket{-,\downarrow,\Uparrow,0}$ has the lowest energy expectation value and is energetically separated from $\ket{-,\downarrow,\Downarrow,0}$ by  $E_{\ket{-,\downarrow,\Downarrow,0}}-E_{\ket{-,\downarrow,\Uparrow,0}}=A/4$, such that lowering the nuclear spin state increases the energy given that the electron is in the orbital state $-$ and its spin orientation is $\downarrow$. Therefore, an excitation exchange between the nuclear spin and the microwave resonator photons corresponds to a population transfer between the states $\ket{-,\downarrow,\Uparrow,n+1}$ and $\ket{-,\downarrow,\Downarrow,n}$ which is why, hereafter, we refer to the coupling between these two states as the transverse nuclear spin-photon coupling.  The energy expectation value difference between $\ket{-,\downarrow,\Uparrow,n+1}$ and $\ket{-,\downarrow,\Downarrow,n}$ reads
\begin{align}
	E_{|-,\downarrow,\Uparrow,n+1\rangle}-E_{|-,\downarrow,\Downarrow,n\rangle}=&-\frac{A}{4}+(\omega_c-\omega_d).
\end{align}
This result unveils that choosing the drive frequency 
\begin{align}
    \textcolor{Black}{\omega_d=\omega_c-\frac{A}{4}+\delta\omega_d}
\end{align}
 allows to tune the nuclear spin splitting in resonance \textcolor{Black}{($\delta\omega_d=0$)} or close to resonance \textcolor{Black}{($\delta\omega_d\neq0$)} with the microwave photons. Hence, the drive allows to compensate \textcolor{Black}{($\delta\omega_d=0$) or partially compensate \textcolor{Black}{($\delta\omega_d\neq0$)}}  the otherwise existing large energy mismatch. \textcolor{Black}{ This effect is sketched in Fig.~\ref{fig:leveldiagram}.} 

\begin{figure}
	\centering 
	\includegraphics[width=\columnwidth]{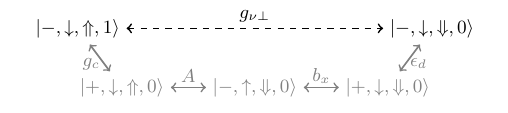}
	\caption{Schematic illustration of the fourth order coupling mechanism underlying the effective transverse nuclear spin-photon coupling $g_{\nu\perp}$ (black dashed line). The inspection of $V^{\mathrm{ R}}$ \eqref{eq:VR} shows that the effective coupling between the two states in black arises due to a fourth order process (gray) including the dipole coupling ($\propto g_c$), the hyperfine interaction ($\propto A$), the magnetic field gradient ($\propto b_x$) and the drive ($\propto\epsilon_d$).}
	\label{fig:coupling_scheme}
\end{figure}

It remains to discuss whether there is an effective nuclear spin photon coupling. For this purpose we consider the QDD system prepared in its ground state and the microwave resonator populated by one photon, i.e. $\ket{-,\downarrow,\Uparrow,1}$.  The analysis of $V^{\mathrm{ R}}$ \eqref{eq:VR} unveils that a fourth order process involving the dipole coupling ($\propto g_c$), the hyperfine interaction ($\propto A$), the magnetic field gradient ($\propto b_x$) and the drive ($\propto\epsilon_d$) couples this state to the state with the cavity photon annihilated and the nuclear spin flipped, as illustrated in Fig.~\ref{fig:coupling_scheme}, such that an effective transverse nuclear spin-photon coupling is expected to emerge. Exploiting the nuclear-spin photon coupling for quantum information applications is mostly of interest if the two states with opposite nuclear spin orientation defining a nuclear spin qubit correspond to the two lowest energy states of the QDD system with and without the drive. While this condition is usually fulfilled in the non-driven scenario,  according to \eqref{eq:Hd} ${2t_c-\omega_d,\,B_z-\omega_d>A/4}$ is required in the presence of the drive and, therefore, the following discussion is restricted to this regime.  

In the next step we derive an effective Hamiltonian for the low-energy nuclear spin subspace interacting with the microwave resonator mode and thereby provide a more formal  derivation of the nuclear spin-photon coupling  including an analytical expression for the coupling strength. For this purpose we use a Schrieffer-Wolff transformation that, in general, decouples the dynamics of two subspaces defined by the projection operators $P_0$ and $Q_0=1-P_0$ \cite{bravyi2011} and results in a block diagonal effective Hamiltonian $H_{\mathrm{eff}}=e^{S}He^{-S}$, with the block off-diagonal and anti-hermitian generator of the transformation $S$. In most cases the transformation $e^{S}$ and $H_{\mathrm{eff}}$ are determined perturbatively \cite{bravyi2011,winkler2003}, as outlined in Appendix~\ref{app:eff_nuclear_spin_photon_coupling}.

Here we are interested in the dynamics of the nuclear spin interacting with the microwave resonator photons while the electron is in the orbital state $-$ and its spin state is $\downarrow$, i.e. the subspace defined by 
\begin{align}
	P_0&=\frac{1-\sigma_z}{2}\frac{1-\tau_z}{2}\nonumber\\
	&=\sum_{n=0}^{\infty}\left(\ket{-\downarrow,\Uparrow,n}\bra{-\downarrow,\Uparrow,n}+\ket{-\downarrow,\Downarrow,n}\bra{-\downarrow,\Downarrow,n}\right).
	\label{eq:P0}
\end{align}
Our  previous reasoning in this section revealed that an effective transverse nuclear spin-photon coupling is expected as a result of a fourth-order process in the perturbation $V^{\mathrm{ R}}$~\eqref{eq:VR}, for which, as discussed in detail in Appendix \ref{app:eff_nuclear_spin_photon_coupling}, we determine the effective Hamiltonian for the subspace set by $P_0$ up to fourth order in the perturbative expansion,
\begin{align}
	H^{\mathrm{R}}_{\nu-\mathrm{ph}}
	=&-\frac{E_{\nu}}{2}\nu_z+\left(\tilde{\omega}_c+\chi\nu_z\right) a^{\dagger}a+K\left(a^{\dagger}a\right)^2\nonumber\\
	&+g_{\nu\perp}\left(\nu_- a+\nu_+ a^{\dagger}\right)+g_{\nu,\mathrm{off}}\left(\nu_+ a+\nu_- a^{\dagger}\right)\nonumber\\
	&+g_{\nu\parallel}\nu_z\left(a+a^{\dagger}\right)\nonumber\\
	&+\left(\tilde{b}_x  +\chi_x  a^{\dagger}a\right)\nu_x+\alpha_9 \left(a+a^{\dagger}\right)\nonumber\\
	&+\alpha_{11} \left(a^2+(a^{\dagger})^2\right)+\alpha_{12}\left(a^{\dagger}a^2+\left(a^{\dagger}\right)^2a\right), 
	\label{eq:Heffsingle}
\end{align}
where $g_{\nu\perp}$ and $g_{\nu\parallel}$ denote the effective transverse and longitudinal nuclear- spin-photon coupling strength and $K$ is the amplitude of a resulting Kerr nonlinearity.
The analytical expression for the coefficients up to fourth order in the perturbation are provided in Appendix~\ref{app:eff_nuclear_spin_photon_coupling}. The diagonal terms collected in the first line differ from $H^{\mathrm{ R}}_0$ due to perturbative corrections manifested in the corresponding coefficients,
\begin{align}
	E_{\nu}&=\frac{A}{4}+\mathcal{O}\left[\left(V^{\mathrm{R}}\right)^4\right],
	\label{eq:Enu}\\
	\widetilde{\omega}_c&=\omega_c-\omega_d-\frac{g_c^2}{2t_c-\omega_c}+\mathcal{O}\left[\left(V^{\mathrm{R}}\right)^4\right],\\
	\chi&=0+\mathcal{O}\left[\left(V^{\mathrm{R}}\right)^4\right],\\
	K&=0+\mathcal{O}\left[\left(V^{\mathrm{R}}\right)^4\right]
	\label{eq:alpha4}.
\end{align}
The above expressions show that only the microwave resonator frequency experiences a correction second order in $V^{\mathrm{R}}$, while the lowest perturbative contribution to the remaining three is $\mathcal{O}\left[\left(V^{\mathrm{R}}\right)^4\right]$. Hence, we have ${E_{\nu},\widetilde{\omega}_c\gg |\chi|,|K|}$ if the system is operated in the regime where the transverse nuclear spin-photon coupling is close to resonance, i.e. $E_{\nu}\approx\widetilde{\omega}_c$.  

The first term in the second line of \eqref{eq:Heffsingle} describes the transverse nuclear spin-photon coupling with coupling strength 
\begin{align}
	g_{\nu\perp}=
	&\frac{1}{2} g_c \frac{A}{4}\frac{ b_x}{2}\frac{\epsilon_d}{4}\times\nonumber\\
	&	\times\left(\frac{1}{ \left(\frac{A}{4}+2t_c-\omega_c\right)(2t_c-\omega_c)(B_z-\omega_c)}\right.\nonumber\\
	&\left.+\frac{1}{ (2t_c-\omega_d)\left(\frac{A}{4}-B_z+\omega_d\right)\left(\frac{A}{4}-2t_c+\omega_d\right)}\right). 
	\label{eq:gnuperp}
\end{align}
We point out that $g_{\nu\perp}\propto g_c A b_x \epsilon_d$ and, therefore, the transverse nuclear spin photon coupling indeed emerges due to the fourth order process illustrated in Fig.~\ref{fig:coupling_scheme} and discussed before. 
Using $A, |\omega_c-\omega_d|\ll t_c, \omega_c, |2t_c-\omega_c|$, we can approximate \eqref{eq:gnuperp} as
\begin{align}
    	g_{\nu\perp} \approx
	\frac{ g_c A b_x \epsilon_d}{32(2t_c-\omega_c)^2(B_z-\omega_c)}.
\end{align}

\begin{figure}
	\centering 
	\includegraphics[width=\columnwidth]{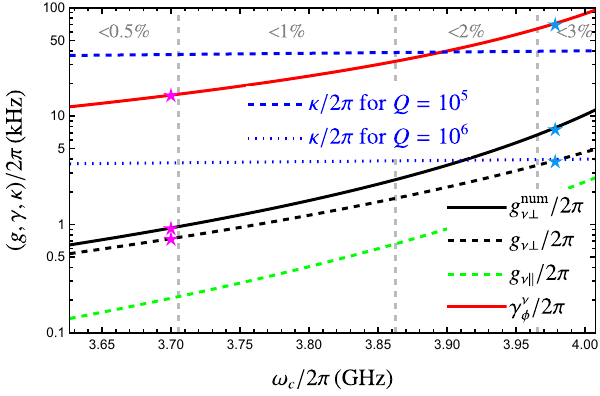}
	\caption{Effective transverse nuclear-spin photon coupling strength $g_{\nu\perp}$ as a function of the microwave resonator frequency $\omega_c$. The dashed black curve corresponds to the derived expression for the coupling strength \eqref{eq:gnuperp}, while the solid black line gives the numerically obtained values for comparison. \textcolor{Black}{The green dashed line gives the longitudinal nuclear spin-photon coupling obtained from \eqref{eq:gnuparallel}.}
		The two blue lines show the microwave resonator photon decay rate $\kappa$ for different resonator quality factors $Q$, while the red line indicates an estimate of the nuclear spin decoherence rate due to hybridization with the charge and spin degree of freedom \textcolor{Black}{obtained as explained in detail in Appendix~\ref{app:eff_decoherence_single_QDD} for the charge (electron spin) decay time $T_1^{\tau}$ $(T_1^{\sigma})$ and decoherence time $T_2^{\tau}$ $(T_2^{\sigma})$ listed in Appendix~\ref{app:eff_decoherence_single_QDD}}. The dashed vertical lines separate parameter domains with increasing degree of hybridization with excited spin, charge and photon states (see text for details). From left to right the degree of hybridization is smaller than  $0.5\%$, $1\%$, $2\%$ and $3\%$. \textcolor{Black}{ The magenta and blue stars indicate the nuclear spin-photon coupling strength and the nuclear spin decoherence rate at two different values of $\omega_c$ for which the dependency of the nuclear spin decoherence rate on the ratio $T_1^{\tau}/T_2^{\tau}$ is presented in Fig.~\ref{fig:nuclear_spin_photon_coupling_ratioT1T2}a and \ref{fig:nuclear_spin_photon_coupling_ratioT1T2}b, respectively. }
		The remaining parameters are ${2t_c=B_z=18\,\upmu\mathrm{eV}}$, ${b_x=1.62\,\upmu\mathrm{eV}}$, ${A=25\,\mathrm{MHz}}$ and  ${g_c/2\pi=(\epsilon_d/2\pi)/4=13\,\mathrm{MHz}}$.}
	\label{fig:nuclear_spin_photon_coupling}
\end{figure}
\begin{figure}
    \centering
    \includegraphics[width=\columnwidth]{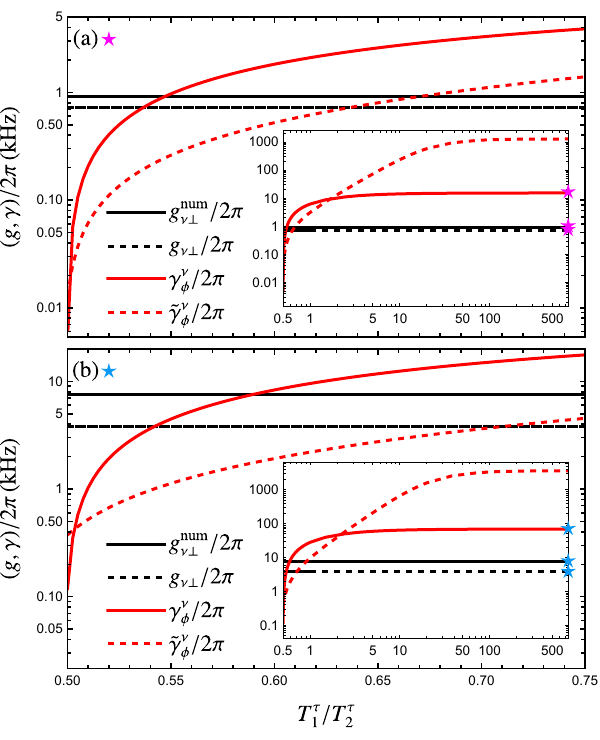}
    \caption{\textcolor{Black}{Nuclear spin decoherence rate $\gamma_{\phi}^{\nu}$ as a function of $T_1^{\tau}/T_2^{\tau}$ for the values of $\omega_c$ indicated by the magenta (a) and blue (b) stars in Fig.~\ref{fig:nuclear_spin_photon_coupling}. The points in the insets highlighted with star symbols and the equally highlighted points in Fig.~\ref{fig:nuclear_spin_photon_coupling} agree. For both the solid and the dashed red line, $T_2^{\tau}$ is kept constant while $T_1^{\tau}$ is varied up until the limiting scenario $T_{2}^{\tau}=2T_{1}^{\tau}$ is approached. In case of the solid red line ($\gamma_{\nu}^{\phi}$) the value of $T_2^{\tau}$ given in the list in Appendix~\ref{app:eff_decoherence_single_QDD} is assumed, while a hundred times shorter $T_2^{\tau}$ time is chosen for the red dashed line  ($\tilde{\gamma}_{\phi}^{\nu}$). The remaining system parameters are as in Fig.~\ref{fig:nuclear_spin_photon_coupling}. }}
    \label{fig:nuclear_spin_photon_coupling_ratioT1T2}
\end{figure}

The most obvious potential application of the effective transverse nuclear spin-photon coupling is the coherent excitation exchange between a nuclear spin qubit and microwave resonator photons. However, this requires strong coupling, i.e. $g_{\nu,\perp}>\kappa,\gamma^{\nu}_{\phi}$, where $\kappa$ is the microwave resonator photon decay rate and $\gamma_{\phi}^{\nu}$ is the nuclear spin qubit decoherence rate. In order to assess the possibility to reach the strong coupling regime, we calculate the coupling strength using system parameters reported for state-of-the-art devices. Figure~\ref{fig:nuclear_spin_photon_coupling} shows a comparison of the effective nuclear spin-photon coupling strength between the analytical result $g_{\nu\perp}$ \eqref{eq:gnuperp} and the result $g_{\nu\perp}^{\mathrm{num}}$ obtained by numerically inspecting the width of the avoided level crossing between the second and the third energy level of $H^{\mathrm{R}}$ \eqref{eq:HR}. This analysis presumes that in the immediate vicinity of the avoided crossing only the coupling between the two involved levels matters such that the width of the avoided crossing corresponds to two times the mutual coupling as predicted by a simple two level model. Both, the analytical and the numerical result show a similar behavior in the considered regime but the analytical result underestimates the coupling strength. 

To gain a better understanding of the origin of the deviation of our analytical result from the numerical one that increases with increasing $\omega_c$, it is insightful to look at the composition of the numerically obtained eigenstates ($\Psi_2$ and $\Psi_3$) corresponding to the energy levels showing the avoided crossing. Thereby we pay particular attention to the combined contribution of the states $\ket{-,\downarrow,\Uparrow,1}$ and $\ket{-,\downarrow,\Downarrow,0}$ to both $\Psi_2$ and $\Psi_3$, i.e.  ${P^{\mathrm{c}}_i=|\braket{-,\downarrow,\Uparrow,1}{\Psi_i}|^2+|\braket{-,\downarrow,\Downarrow,0}{\Psi_i}|^2},\,i=2,3$, at the point of the avoided crossing. It turns out that $1-P^{\mathrm{c}}_i$ increases with increasing $\omega_c$ for the parameter regime considered in Fig.~\ref{fig:nuclear_spin_photon_coupling}, as highlighted by the gray dashed lines in the figure that separate parameter domains corresponding to different ranges of $1-\min(P^{\mathrm{c}}_2,P^{\mathrm{c}}_3)$. This observation explains the difference between the analytical and the numerical result for the coupling strength: a deviation of $1-\min(P^{\mathrm{c}}_2,P^{\mathrm{c}}_3)$ from 0 implies that the states undergoing the avoided crossing are not purely given by a hybridization of the unperturbed states  $\ket{-,\downarrow,\Uparrow,1}$ and $\ket{-,\downarrow,\Downarrow,0}$ but also the excited orbital ($+$) and electron spin ($\uparrow$) states as well as states differing from the unperturbed states with respect to the photon number contribute increasingly, which is why we refer to the quantity $1-\min(P^{\mathrm{c}}_2,P^{\mathrm{c}}_3)$ as the degree of hybridization. Since we use a perturbative approach for the SW transformation, $g_{\nu\perp}$ and $g_{\nu\perp}^{\mathrm{num}}$ show the best agreement for parameter regimes within which the degree of hybridization, $1-\min(P^{\mathrm{c}}_2,P^{\mathrm{c}}_3)$, is close to $0$. In general, the accuracy of $g_{\nu\perp}$ can be increased by including higher order corrections in the perturbative determination of $H_{\mathrm{eff}}$ \eqref{eq:Heffexpansion} at the cost of additional terms and more complex expressions for the appearing coefficients which is why our discussion is limited to fourth order corrections in $V^{\mathrm{R}}$ \eqref{eq:VR}. Nevertheless, the analytical result  $g_{\nu\perp}$ allows to estimate the order of magnitude of the effective nuclear spin-photon coupling strength reliably. 

To assess whether the strong nuclear spin-photon coupling regime is in reach or not, it remains to compare the coupling strength to the microwave resonator decay rate $\kappa$ and the nuclear spin qubit state decoherence rate $\gamma^{\nu}_{\phi}$. The blue curves in Fig.~\ref{fig:nuclear_spin_photon_coupling} give the microwave resonator photon decay rate $\kappa$ for $Q=10^5$ and $Q=10^6$, where $Q=\omega_c/\kappa$ is the resonator quality factor.  While the reported quality factors $\mathcal{Q}\gtrsim10^6$ require architectures with strong magnetic shielding \cite{megrant2012,bruno2015}, quality factors $Q\approx10^5$ are available for setups, as the one considered in this article, with substantial magnetic fields \cite{samkharadze2016,kroll2019}. Figure~\ref{fig:nuclear_spin_photon_coupling} shows that $\kappa>g_{\nu\perp}$ for $\mathcal{Q}=10^5$ in the considered parameter domain. Therefore, the strong coupling condition is violated and we see  that it requires to increase the quality factor by at least one order of magnitude to bring $\kappa$ in accordance with the strong coupling condition. 

In addition we also check the compatibility of the nuclear spin decoherence rate with strong coupling. In order to do so, the decoherence rate between the nuclear spin qubit states due to charge decay and dephasing as well as electron spin decay and dephasing is numerically calculated following the procedure outlined in detail in Appendix~\ref{app:nuclear_spin_decoherence_rate}. Assuming a state-of-the-art decay time $T_{1}^{\tau}$ $(T_{1}^{\sigma})$ and decoherence time $T_{2}^{\tau}$ $(T_{2}^{\sigma})$ for the charge (electron spin), we find the nuclear spin decoherence rate $\gamma^{\nu}_{\phi}$ given by the red curve in Fig.~\ref{fig:nuclear_spin_photon_coupling}. Obviously, $\gamma^{\nu}_{\phi}>g_{\nu\perp}$ such that the strong coupling condition is again violated. 

\textcolor{Black}{However, we find that shorter charge decay times, i.e. a faster relaxation of the excited charge qubit state to its ground state at a constant charge decoherence time lead to an increased nuclear spin decohrence time, as shown in Fig.~\ref{fig:nuclear_spin_photon_coupling_ratioT1T2}. The figure unveils that, for the considered system parameters, $g_{\nu\perp}>\gamma_{\phi}^{\nu}$ can be realized close to the limiting scenario $T_2^{\tau}=2T_{1}^{\tau}$. Interestingly, $g_{\nu\perp}>\gamma_{\phi}^{\nu}$ is still observed close to $T_2^{\tau}=2T_{1}^{\tau}$ if the charge decoherence time $T_{2}^{\tau}$ is two orders of magnitude shorter (red dashed lines in Fig.~\ref{fig:nuclear_spin_photon_coupling_ratioT1T2}) than the longest reported ones (red solid lines in Fig.~\ref{fig:nuclear_spin_photon_coupling_ratioT1T2}).}

\textcolor{Black}{Finally, we briefly inspect the longitudinal nuclear spin-photon coupling, because longitudinal coupling has potential use for qubit readout \cite{didier2015} as well as the implementation two qubit gates \cite{jin2012,royer2017,harvey2018} and is actively investigated experimentally \cite{bottcher2022}. Fig.~\ref{fig:nuclear_spin_photon_coupling} shows the longitudinal coupling strength $g_{\nu\parallel}$ obtained from Eq.~\eqref{eq:gnuparallel}. We find that the transverse $g_{\nu\perp}$ and the longitudinal coupling strength $g_{\nu\parallel}$ are of the same order of magnitude, while $g_{\nu\parallel}<g_{\nu\perp}$. Within the scope of this work, the potential usage of the longitudinal coupling is not further investigated.}

In summary, for state-of-the-art architectures the strong nuclear spin-photon coupling is not in reach because $\kappa$ exceeds the effective nuclear spin-photon coupling strength that we predict to be in the kHz domain. Therefore coherent excitation exchange between a nuclear spin qubit and microwave resonator photons is not possible. Nevertheless, the derived transverse nuclear spin-photon coupling can be exploited for quantum information applications as demonstrated in the next section.



\section{Dispersive nuclear spin gate \label{sec:dispersive_nuclear_spin_gate}}

	\begin{figure*}
	\centering
	\includegraphics[width=0.8\textwidth]{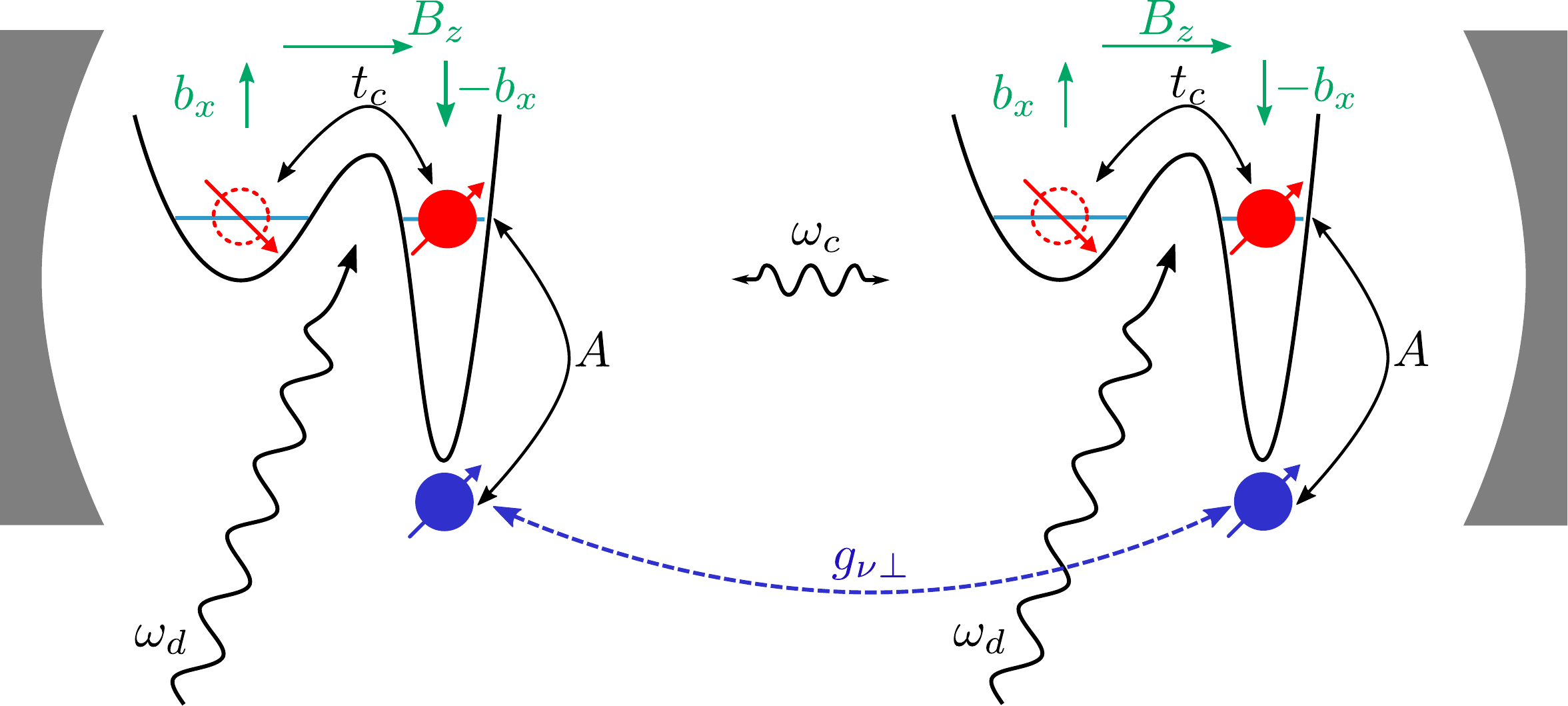}
	\caption{Schematic illustration of two driven, unbiased ($\epsilon=0$) QDD systems coupled to a microwave resonator mode. The individual QDD systems are similar to the one presented in Fig.~\ref{fig:quantumdot-donor system}.  The resonator mediates an interaction between the two nuclear spins of the two distant QDD systems if the effective coupling between the individual nuclear spins and the resonator photons is tuned to the dispersive regime. A detailed discussion of the explicit form of the interaction strength $g_{\nu\perp}$ between the nuclear spins and its potential for quantum information applications is provided in Section \ref{sec:dispersive_nuclear_spin_gate}.}
	\label{fig:driven_donor_coupling}
\end{figure*}
As shown in the previous section, the transverse nuclear spin-photon coupling is too weak to allow for coherent excitation exchange between a nuclear spin qubit and the microwave resonator mode on resonance (in the rotating reference frame). However, in this section we explore a further possibility to make use of the effective nuclear spin-photon coupling for a quantum information application.  Motivated by the theoretical prediction of a microwave resonator photon mediated quantum gate between two electron spin qubits dispersively coupled to the resonator \cite{benito2019a}, we investigate whether a similar interaction can be realized between two nuclear spin qubits. The mentioned dispersive coupling describes a regime in which the two electron spin qubits are coupled to the resonator but the qubit frequency and the resonator mode are off-resonant such that the coherent excitation exchange between the qubits and the resonator mode is strongly suppressed. \textcolor{Black}{This in turn implies that the resonator is only virtually populated and, therefore, the photon decay has little impact on the system, while, at the same time, the system can be operated in a regime with $g_{\nu\perp}>\gamma_{\phi}^{\nu}$ by approaching the limiting scenario $T_{2}^{\tau}=2T_{1}^{\tau}$.} 

First, the model of the driven QDD system introduced in Sec.~\ref{sec:model} has to be extended. For simplicity, we assume that two identical driven non-detuned QDD systems are coupled to a common microwave resonator as depicted in Fig.~\ref{fig:driven_donor_coupling}.  The Hamiltonian of this system can be written as
\begin{align}
	\hat{H}(t)=\sum_{i=1}^2 \left(H_{\mathrm{QDD}}^{(i)}+	H_{\mathrm{int}}^{(i)}+H_{d}^{(i)}(t)\right)+H_{\mathrm{cav}},
	\label{eq:Hhat}
\end{align}
where the index $i=1,2$ labels the two QDD systems and the labeled terms are obtained from \eqref{eq:HQDD}-\eqref{eq:Hd} by the replacements $\tau_{\beta}\rightarrow \tau_{\beta}^{(i)}$, $\sigma_{\beta}\rightarrow \sigma_{\beta}^{(i)}$ and $\nu_{\beta}\rightarrow \nu_{\beta}^{(i)}$ with $\beta\in\left\{x,y,z,+,-\right\}$. Starting from this Hamiltonian we follow the procedure that allowed us to derive the effective nuclear spin-photon coupling. We begin by transforming $\hat{H}(t)$ to a rotating reference frame defined by 
\begin{align}
	\hat{U}_{\mathrm{R}}(t)=U_{\mathrm{ R,sys}}^{(1)}(t)U_{\mathrm{ R,sys}}^{(2)}(t)U_{\mathrm{R,a}}(t)
	\label{eq:URhat}
\end{align}
before applying a Schrieffer-Wolff transformation that yields the effective dynamics of the two nuclear spins interacting with the resonator photons, i.e. the subspace defined by the projection operator 
\begin{align}
\hat{P}_{0}=\frac{1-\sigma_z^{(1)}}{2}	\frac{1-\sigma_z^{(2)}}{2}	\frac{1-\tau_z^{(1)}}{2}	\frac{1-\tau_z^{(2)}}{2}.
\label{eq:P0hat}
\end{align}
Again, the effective Hamiltonian is determined up to fourth order in the perturbative expansion to capture the fourth order process underlying the transverse nuclear spin-photon coupling. According to the more detailed derivation presented in Appendix~\ref{app:coupled_nuclear_spins}, the effective Hamiltonian reads 
\begin{align}
	\hat{H}^{\mathrm{R}}_{\nu-\mathrm{ph}}
	=&\sum_{i=1}^{2}\left[-\frac{E_{\nu}}{2}\nu_z^{(i)}+\chi\nu_z^{(i)} a^{\dagger}a\right.\nonumber\\
	&\left.\quad\quad+g_{\nu\perp}\left(\nu_-^{(i)} a+\nu_+^{(i)} a^{\dagger}\right)\right.\nonumber\\
	&\quad\quad+g_{\nu,\mathrm{off}}\left(\nu_+^{(i)} a+\nu_-^{(i)} a^{\dagger}\right)+g_{\nu\parallel}\nu_z^{(i)}\left(a+a^{\dagger}\right)\nonumber\\
	&\quad\quad\left.+\left(\tilde{b}_x  +\chi_x  a^{\dagger}a\right)\nu_x^{(i)}\right]\nonumber\\
	&+\bar{\omega}_c a^{\dagger}a+\bar{K}\left(a^{\dagger}a\right)^2+\bar{\alpha}_9 \left(a+a^{\dagger}\right)\nonumber\\
	&+\bar{\alpha}_{11} \left(a^2+(a^{\dagger})^2\right)+\bar{\alpha}_{12}\left(a^{\dagger}a^2+\left(a^{\dagger}\right)^2a\right).
	\label{eq:Heffhat}
\end{align}
A comparison with \eqref{eq:Heffsingle} shows a similar structure. However, the coefficients marked by a bar include additional perturbative corrections compared to the ones without the bar in \eqref{eq:Heffsingle}. These additional corrections are given in Appendix~\ref{app:coupled_nuclear_spins} and originate from processes that involve both QDD systems and are therefore not present if only one such system interacts with the resonator mode. 

\begin{figure}
	\centering
	\includegraphics[width=\columnwidth]{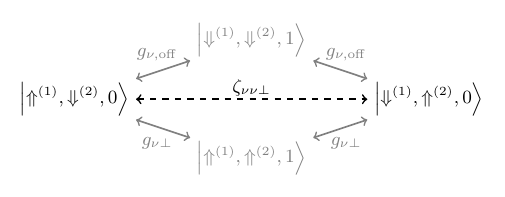}
\caption{Schematic illustration of the second order coupling mechanisms underlying the effective coupling $\zeta_{\nu\nu\perp}$ between the nuclear spins of the two QDD systems (1) and (2) (black dashed line). The inspection of $\hat{H}_{\mathrm{eff}}$~\eqref{eq:Heffhat} shows that there are two possible second order processes (gray) that couple the two states in black.}
\label{fig:effectivecouplingmechanismnuclearspins}
\end{figure}

Given two nuclear spins that interact with the same mode of a microwave resonator as described by the above Hamiltonian, it seems apparent that the resonator can mediate an interaction between the nuclear spins. In order to develop an intuition for a possible resulting interaction, we consider the system configuration with the two nuclear spins anti-aligned and the cavity being empty, i.e. the system is prepared in the state $\ket{\Uparrow^{(1)},\Downarrow^{(2)},0}$, where the first two entries represent the nuclear spin state of  QDD system 1 and 2 while the last entry gives the photon occupation number of the resonator mode. A closer look at $\hat{H}^{\mathrm{ R}}_{\nu-\mathrm{ph}}$~\eqref{eq:Heffhat} unveils that the two second order processes sketched in Fig.~\ref{fig:effectivecouplingmechanismnuclearspins} connect this state to the state with both nuclear spins flipped. In both cases the intermediate state includes the occupation of the resonator mode with a single photon. However, if the system is operated in the dispersive nuclear spin-photon coupling regime characterized by a detuning between the cavity mode and the nuclear spin splitting in the rotating reference frame, the intermediate states are only virtually populated and the resonator is expected to mediate an interaction between the nuclear spins without a real intermediate population of the resonator mode. 

Motivated by this line of reasoning, we aim at deriving an effective Hamiltonian that captures the interaction between the nuclear spins of the two QDD systems mediated by an empty microwave resonator. Starting from $\hat{H}^{\mathrm{ R}}_{\nu-\mathrm{ph}}~$\eqref{eq:Heffhat}, this means that we are interested in the effective dynamics of the subspace corresponding to the projection operator
\begin{align}
\hat{P}_{0}^{\nu-\nu}=\ket{0}\bra{0},
\end{align}
with $\ket{0}$ the resonator vacuum state.  Another Schrieffer-Wolff transformation up to second order in the perturbative expansion (see Appendix~\ref{app:coupled_nuclear_spins} for more details) yields
\begin{align}
	\hat{H}^{\mathrm{R}}_{\nu-\nu}=&-\sum_{i=1}^{2}\frac{\bar{E}_{\nu}}{2}\,\nu_z^{(i)}+\delta \bar{E}_{\nu}\,\nu_z^{(1)}\nu_z^{(2)}\nonumber\\
	&+\sum_{i=1}^2\left(\bar{b}_x+\sum_{j\neq i}c_2\nu_z^{(j)} \right)\nu_x^{(i)}\nonumber\\
	&+\zeta_{\nu\nu\perp}\left(\nu_+^{(1)}\nu_-^{(2)}+\nu_-^{(1)}\nu_+^{(2)}\right)\nonumber\\
	&+\zeta_{\nu\nu\parallel}\left(\nu_+^{(1)}\nu_+^{(2)}+\nu_-^{(1)}\nu_-^{(2)}\right),
	\label{eq:HhatRnunu}
\end{align}
with the explicit expressions for all the coefficients listed in Appendix~\ref{app:coupled_nuclear_spins}. We point put that the interaction term in the third line agrees with our expectation, and, in particular, that the corresponding coupling strength
\begin{align}
	\zeta_{\nu\nu\perp}=\frac{g_{\nu\perp}^2}{E_{\nu}-2\chi-\bar{\omega}_c-\bar{K}}-\frac{g_{\nu,\mathrm{off}}^2}{E_{\nu}-2\chi+\bar{\omega}_c+\bar{K}},
	\label{eq:zetanunuperp}
\end{align}
 reflects the two paths shown in Fig.~\ref{fig:effectivecouplingmechanismnuclearspins}. 
 
 Next, we transform $\hat{H}^{\mathrm{R}}_{\nu-\nu}$~\eqref{eq:HhatRnunu} to the rotating reference frame set by the unitary 
\begin{align}
	\hat{U}_{\mathrm{ R1}}=\exp\left[i \left(-\sum_{i=1}^2\frac{\bar{E}_{\nu}}{2}\nu_z^{(i)}\right) t\right]. 
\end{align}
Additionally applying the rotating wave approximation to the transformed Hamiltonian, the non-resonant terms can be neglected and we finally arrive at 
\begin{align}
	\hat{H}^{\mathrm{R,R1}}_{\nu-\nu}=\delta \bar{E}_{\nu}\,\nu_z^{(1)}\nu_z^{(2)}+\zeta_{\nu\nu\perp}\left(\nu_+^{(1)}\nu_-^{(2)}+\nu_-^{(1)}\nu_+^{(2)}\right). 
\end{align}
Equipped with the above Hamiltonian it is straightforward to calculate the corresponding time evolution operator up to a global phase 
\begin{align}
	\hat{U}^{\mathrm{ R,R1}}_{\nu-\nu}(t,0)=&\begin{pmatrix}
		1 & 0 & 0 & 0 \\
		0 & \cos\left(\zeta_{\nu\nu\perp}\,t\right) & -i \sin\left(\zeta_{\nu\nu\perp}\,t\right) & 0  \\
		0&  -i \sin\left(\zeta_{\nu\nu\perp}\,t\right)  & \cos\left(\zeta_{\nu\nu\perp}\,t\right) & 0 \\
		0 & 0 & 0 & 1
	\end{pmatrix}\nonumber\\
	&\times \begin{pmatrix}
	e^{-i 2\delta\bar{E}_{\nu} t} & 0 & 0 & 0 \\
	0 & 1 & 0 & 0 \\
	0 & 0 & 1 & 0 \\
	0 & 0 & 0 & e^{-i 2\delta\bar{E}_{\nu} t}
	\end{pmatrix},
	\label{eq:time_evolution}
\end{align}
with the matrix representation chosen according to the basis 
$\{\ket{\Uparrow^{(1)},\Uparrow^{(2)}},\ket{\Uparrow^{(1)},\Downarrow^{(2)}},\ket{\Downarrow^{(1)},\Uparrow^{(2)}},\ket{\Downarrow^{(1)},\Downarrow^{(2)}}\}$.

We note that the above time evolution corresponds to the product of the first matrix in Eq.~\eqref{eq:time_evolution} with a controlled phase gate (CPHASE) up to single qubit rotations. Besides that, the evolution described by the two matrices in Eq.~\eqref{eq:time_evolution} is governed by two different timescales characterized by $\zeta_{\nu\nu\perp}$ and $\delta\bar{E}_{\nu}$, respectively. In order to assess the different timescales, we first recall that ${E_{\nu},\bar{\omega_c}\gg\bar{K},\chi,\bar{\alpha}_9,\bar{\alpha}_{11},g_{\nu\perp},g_{\nu\mathrm{off}},g_{\nu\parallel}}$ because all the coefficients on the right hand side of the inequality are determined by fourth order correction terms, while the dominant contributions to the coefficients on the left hand side are non-perturbative (see the explicit form of all the coefficients presented in Appendix~\ref{app:coupled_nuclear_spins}). Moreover, the nuclear spins are dispersively coupled to the microwave resonator photons, i.e. ${|g_{\nu,\perp}/(E_{\nu}-2\chi-\bar{\omega}_c-\bar{K})|\approx |g_{\nu,\perp}/(E_{\nu}-\bar{\omega}_c)|\ll 1}$. Without violating this condition the system can be tuned to a regime with $E_{\nu},\bar{\omega}_c>|E_{\nu}-\bar{\omega}_c|$ thereby ensuring a non-negligible nuclear spin-nuclear spin coupling $\zeta_{\nu\nu\perp}$ \eqref{eq:zetanunuperp}. In this operating regime the first term in \eqref{eq:zetanunuperp} gives the dominant contribution to $\zeta_{\nu\nu\perp}$. Looking at all the terms contributing to $\delta\bar{E}_{\nu}$ \eqref{eq:deltaEbarnu} individually, the two terms appearing in the third line of Eq.~\eqref{eq:deltaEbarnu} seem to be the dominant ones both with an absolute value being approximately $1/2$ of the absolute value of the dominant contribution to $\zeta_{\nu\nu\perp}$, while the absolute values \textcolor{Black}{of} all other terms are small compared to the two mentioned ones. However, the two terms appear with opposite signs and the non-zero absolute value of their difference $|g_{\nu\perp}^2/(\bar{K}-E_{\nu}+\bar{\omega}_c+2\chi)\cdot \chi/(\bar{K}-E_{\nu}+\bar{\omega}_c)|$ is small compared to the dominant contribution to $\zeta_{\nu\nu\perp}$ because $|\chi/(\bar{K}-E_{\nu}+\bar{\omega}_c)|\ll1$ and, therefore, $\zeta_{\nu\nu\perp}\gg\delta\bar{E}_{\nu}$. This immediately implies that the time evolution generated by the second matrix in \eqref{eq:time_evolution} on times scales with $\delta\bar{E}_{\nu}t\ll 1$ is negligible compared to the first matrix. 

Taking the above considerations into account it is easy to verify that  the time evolution realizes a $\sqrt{i\mathrm{SWAP}}$ gate when setting the evolution time to  $t_{\sqrt{i\mathrm{SWAP}}}=\pi /(4\zeta_{\nu\nu\perp})$,
\begin{align}
	\hat{U}^{\mathrm{ R,R1}}_{\nu-\nu}\left(\pi /(4\zeta_{\nu\nu\perp}),0\right)=\begin{pmatrix}
		1 & 0 & 0 & 0 \\
		0 & 1/\sqrt{2} & -i /\sqrt{2}  & 0  \\
		0&  -i /\sqrt{2}   & 1/\sqrt{2} & 0 \\
		0 & 0 & 0 & 1 
	\end{pmatrix}. 
\end{align}
Evolving for twice the time  $t_{\sqrt{i\mathrm{SWAP}}}$ yields an $i\mathrm{SWAP}$ gate ($t_{i\mathrm{SWAP}}=\pi /(2\zeta_{\nu\nu\perp})$):
 \begin{align}
	\hat{U}^{\mathrm{ R,R1}}_{\nu-\nu}\left(\pi /(2\zeta_{\nu\nu\perp}),0\right)=\begin{pmatrix}
		1 & 0 & 0 & 0 \\
		0 & 0 & -i  & 0  \\
		0&  -i   & 0& 0 \\
		0 & 0 & 0 & 1 
	\end{pmatrix}.
\end{align}

The theoretical prediction of the two possible quantum gates assuming fully coherent evolution is a promising result but in practice, it matters whether the gates can be realized in an actual device in the presence of decoherence. In order to answer this question we run numerical simulations accounting for charge decay and dephasing, electron spin decay and dephasing, as well as microwave resonator photon decay\textcolor{Black}{, while the intrinsic nuclear spin decay and dephasing is not included because the corresponding timescale \cite{steger2012,pla2013,saeedi2013,muhonen2014} are orders of magnitude longer than those reported for the aforementioned decoherence processes}. Solving the Lindblad master equation for the full system including the two QDD systems coupled to the microwave resonator would require large computational resources due to the large Hilbert space. To circumvent this problem we developed an effective decoherence model. The detailed description of the construction of the model is provided in Appendix~\ref{app:effective decoherence model} and can be summarized as follows: First, the master equation for a single QDD system coupled to the microwave resonator is solved for initial density matrices restricted to the nuclear spin subspace with zero resonator photons. Then effective decoherence rates assuming a three level model including the two nuclear spin qubit states and a leakage state are extracted from these results. We demonstrate that an effective Hamiltonian describing the interaction of the nuclear spin of the QDD system with the microwave resonator mode combined with the effective coherence rates reproduces the decoherence dynamics of the full system with high accuracy on the timescale of interest. This justifies applying the effective decoherence model to the larger setup with the nuclear spin of two QDD system coupled to the microwave resonator. 

Performing parameters scans over parameter domains obeying all the introduced restrictions, we find that for the system parameters mentioned in Appendix~\ref{app:effective decoherence dynamics nuclear spin gate }  average gate fidelities $\bar{F}_{\mathrm{gate}}=0.90$ for the $\sqrt{i\mathrm{SWAP}}$ gate and $\bar{F}_{\mathrm{gate}}=0.80$  for the $\mathrm{SWAP}$ gate are in reach assuming \textcolor{Black}{a cavity quality factor of $Q=10^5$, $T_{1}^{\tau}=T_{2}^{\tau}/2$, as well as the reported decay and decohrence times $T_{2}^{\tau},T_{1}^{\sigma}$ and $T_{2}^{\sigma}$} listed in Appendix~\ref{app:nuclear_spin_decoherence_rate}. The envelope of the numerically calculated average gate fidelity in the rotating reference frame $\mathrm{R}$ (the average fidelity oscillates quickly because we do not apply the subsequent transformation to the rotating reference frame $\mathrm{R1}$) as a function of time for both quantum gates is shown in Fig.~\ref{fig:average_fidelity_effective_model}. The plot shows that the gate times for the $\sqrt{i\mathrm{SWAP}}$ and the $i\mathrm{SWAP}$ gate are $t_{\sqrt{i\mathrm{SWAP}}}=3.44\,\mathrm{ms}$ and $t_{i\mathrm{SWAP}}=7.97\,\mathrm{ms}$, respectively. Moreover, the curves for the coherent evolution approaching a gate fidelity of 1 support the reasoning that the time evolution generated by the second matrix in \eqref{eq:time_evolution} is negligible on the considered timescale.

\textcolor{Black}{The predicted gate times of the order of several milliseconds approve that the intrinsic nuclear spin decay and decoherence characterized by times exceeding tens of seconds \cite{steger2012,pla2013,saeedi2013,muhonen2014} can be neglected in the numerical simulations. }

\textcolor{Black}{Another question that arises due to the long gate times is whether the  results found in this work persist in the presence of the counter rotating terms neglected in the presented analysis by applying the RWA. To address this question, the coherent evolution of the full system composed of two DQD systems coupled to a microwave resonator with and without the RWA applied is compared in Appendix~\ref{app:justification_RWA} and, thereby, we conclude that it is indeed justified to apply the RWA for the set of system parameters for which we find the best average gate fidelities for the $i\mathrm{SWAP}$ and the $\sqrt{i\mathrm{SWAP}}$ gate.}

\begin{figure}
\centering
\includegraphics[width=\columnwidth]{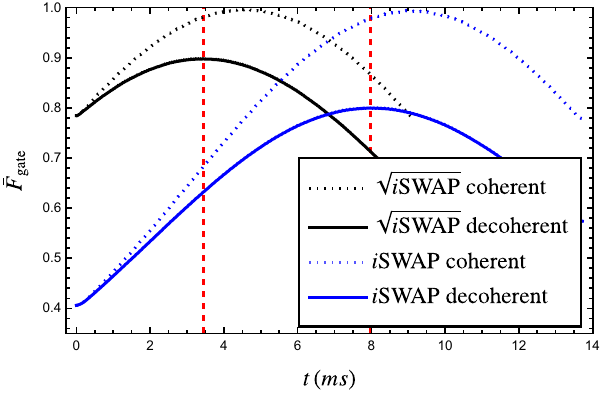}
\caption{Envelope of the average gate fidelity $\bar{F}_{\mathrm{gate}}$ for the $\sqrt{i\mathrm{SWAP}}$ and the $i\mathrm{SWAP}$ gate as a function of time. The dotted lines correspond to the coherent evolution governed by $\hat{H}^{\mathrm{R}}=\hat{U}_{\mathrm{R}}(t)\hat{H}(t)\hat{U}^{\dagger}_{\mathrm{R}}(t)+i\dot{\hat{U}}_{\mathrm{R}}(t)\hat{U}^{\dagger}_{\mathrm{R}}(t)$ (see Eqs.~\eqref{eq:Hhat} and \eqref{eq:URhat}) in both cases while the solid lines obtained with the effective decoherence model outlined in Appendix~\ref{app:effective decoherence model} include decoherence effects. From left to right, the vertical dashed lines give the optimal gate times extracted from our numerical simulations for the $\sqrt{i\mathrm{SWAP}}$ and the $i\mathrm{SWAP}$ gate. \textcolor{Black}{For the decoherence model used for this plot we assume a cavity quality factor of $Q=10^5$, $T_{1}^{\tau}=T_{2}^{\tau}/2$  and the decay and decoherence times $T_{2}^{\tau},T_{1}^{\sigma}$ and $T_{2}^{\sigma}$ listed in Appendix~\ref{app:nuclear_spin_decoherence_rate}.} The system parameters agree with those given in Appendix~\ref{app:effective decoherence dynamics nuclear spin gate }.}
\label{fig:avg_fidelity_evolution}
\end{figure}

\section{Conclusion \label{sec:conclusion}}
We provided a theoretical study of a hybrid QDD architecture consisting of a gate defined Si quantum dot and a thereto laterally displaced phosphorous donor atom implanted in the Si host material in particular with regard to its potential use for quantum information applications. 
Our analysis showed that periodically driving the QDD detuning allows to compensate the frequency mismatch between the donor nuclear spin-splitting and the microwave resonator frequency. Thus resonant nuclear spin-photon coupling is possible. However, our analysis unveiled that the strong coupling regime is not in reach preventing the coherent excitation exchange between the nuclear spin and the photons. 
Nevertheless, using realistic parameters, we demonstrated that dispersively coupling the nuclear spins of two driven QDD system to a microwave resonator allows the implementation of a nuclear spin $\sqrt{i\mathrm{SWAP}}$ and $i\mathrm{SWAP}$ gate mediated by virtual resonator photons with an average gate fidelity exceeding  $\bar{F}_{\mathrm{gate}}=0.95$ and $\bar{F}_{\mathrm{gate}}=0.90$, respectively. These numbers were obtained from elaborate numerical simulations accounting for charge decay and dephasing, electron spin decay and dephasing, as well as resonator photon decay. 

\section*{Acknowledgments}
This work has been supported by ARO grant number W911NF-15-1-0149.

\appendix
\section{Effective Hamiltonian nuclear spin-photon coupling \label{app:eff_nuclear_spin_photon_coupling}}
The fundamental challenge to perform a Schrieffer Wolff transformation is the determination of the generator $S$ of the unitary transformation $e^S$. It is a reliable strategy to use a perturbative expansion $S=\sum_{n=1}^{\infty}S_n$ where $S_n$ is an operator of $n$-th order in the perturbation. Choosing this ansatz it can be shown that the first three contributions to $S$ have to obey the following relations \cite{bravyi2011}:
\begin{align}
	\label{eq:defS1}
	\left[H^{\mathrm{R}}_{0},S_{1}\right]=& V^{\mathrm{ R}}_{\mathrm{od}},\\
	\label{eq:defS2}
	\left[H^{\mathrm{ R}}_{0},S_{2}\right]=&-\left[V^{\mathrm{R}}_{\mathrm{d}},S_{1}\right],\\
			\label{eq:defS3}
	\left[H^{\mathrm{ R}}_{0},S_{3}\right]=&-\left[V^{R}_{\mathrm{d}},S_{2}\right]+\frac{1}{3}\left[S_{1},\left[S_{1},V^{\mathrm{R}}_{\mathrm{od}}\right]\right],
\end{align}
with 
\begin{align}
	V^{\mathrm{R}}_{\mathrm{d}}=&P_{0}V^{\mathrm{ R}}P_{0}+Q_{0}V^{\mathrm{R}} Q_{0},
\end{align}
and
\begin{align}
	V^{\mathrm{R}}_{\mathrm{od}}=&P_{0}V^{R}Q_{0}+Q_{0}V^{\mathrm{R}} P_{0},
\end{align}
the block diagonal and the block off-diagonal parts of $V^{\mathrm{ R}}$~\eqref{eq:VR}, respectively. The defining conditions \eqref{eq:defS1}~-~\eqref{eq:defS3} together with the known commutation relations for the Pauli operators and the photonic creation and annihilation operators allow the determination of $S_1$-$S_3$, which in turn are sufficient to compute the effective Hamiltonian governing the dynamics of the subspace set by $P_0$ up to fourth order in the perturbation $V^{R} $\cite{bravyi2011}, 
\begin{align}
	H^{\mathrm{R}}_{\nu-\mathrm{ph}}=&	P_{0}	H^{\mathrm{ R}}_{0}P_{0}+V^{\mathrm{R}}P_{0}+\sum_{n=2}^{4}H^{\mathrm{R},(n)}_{\nu-\mathrm{ph}},
	\label{eq:Heffexpansion}
\end{align}
with
\begin{align}
	H^{\mathrm{R},(2)}_{\nu-\mathrm{ph}}=&\frac{1}{2}	P_{0}\left[S_{1},V^{\mathrm{R}}_{\mathrm{od}}\right]	P_{0},\\
	H^{\mathrm{R},(3)}_{\nu-\mathrm{ph}}=&\frac{1}{2}	P_{0}\left[S_{2},V^{\mathrm{R}}_{\mathrm{od}}\right]	P_{0},\\
	H^{\mathrm{R},(4)}_{\nu-\mathrm{ph}}=&\frac{1}{2}	P_{0}\left[S_{3},V^{\mathrm{R}}_{\mathrm{od}}\right]	P_{0}\nonumber\\
	&-\frac{1}{24}	P_{0} \left[S_{1},\left[S_{1},\left[S_{1},V^{\mathrm{R}}_{\mathrm{od}}\right]\right]\right]P_{0}.
\end{align}
The specific form of the effective \textcolor{Black}{Hamiltonian} describing the dynamics of the nuclear spin of the driven QDD system interacting with microwave resonator photons is given by 
\begin{align}
	H^{\mathrm{R}}_{\nu-\mathrm{ph}}
=&-\frac{E_{\nu}}{2}\nu_z+\left(\tilde{\omega}_c+\chi\nu_z\right) a^{\dagger}a+K\left(a^{\dagger}a\right)^2\nonumber\\
&+g_{\nu\perp}\left(\nu_- a+\nu_+ a^{\dagger}\right)+g_{\nu,\mathrm{off}}\left(\nu_+ a+\nu_- a^{\dagger}\right)\nonumber\\
&+g_{\nu\parallel}\nu_z\left(a+a^{\dagger}\right)\nonumber\\
&+\left(\tilde{b}_x  +\chi_x  a^{\dagger}a\right)\nu_x+\alpha_9 \left(a+a^{\dagger}\right)\nonumber\\
&+\alpha_{11} \left(a^2+(a^{\dagger})^2\right)+\alpha_{12}\left(a^{\dagger}a^2+\left(a^{\dagger}\right)^2a\right). 
	\label{eq:Heffappnedix}
\end{align}
We find analytical expressions for all the coefficients appearing in the above Hamiltonian:
\begin{widetext}
	\begingroup
	\allowdisplaybreaks
\begin{align}
E_{\nu}=&\frac{A}{4}+\frac{A b_x^2 \epsilon_d^2}{8 (\textcolor{Black}{2t_c} -\omega_d)^2 \left[A^2-16 (B_z-\textcolor{Black}{\omega _d})^2\right]}+\frac{A^2 \epsilon_d^2}{256 (B_z-\omega_d) (2t_c -\omega_d)^2},
\label{eq:firstcoefficient}\\
		\widetilde{\omega}_c=&\omega_c-\omega_d-\frac{g_c^2}{2t_c -\omega_c}-\frac{g_c^4}{(2t_c -\omega_c)^3}\nonumber\\
		&	+\Bigg [g_c^2 \bigg\{-\left(A^4 [2t_c -\omega_c] [2t_c -\omega _d]^2\right)  \nonumber\\
		&\quad\quad\quad\,
		+4 A^2 \bigg(4 B_z^2 [2t_c -\omega_c]
		[2t_c -\omega_d]^2 
		+B_z \Big\{8 g_c^2 [2t_c -\omega_d]^2-[2t_c -\omega_c] \Big(8 \omega_c [2t_c
		-\omega_d]^2   +\epsilon_d^2 [-4t_c +\omega_c+\omega_d]\Big)\Big\} \nonumber\\
		&\quad\quad\quad\quad\quad\quad\quad-8 g_c^2 \omega_c [2t_c
		-\omega_d]^2
		+\omega_c [2t_c -\omega_c] \Big\{4 \omega_c [2t_c -\omega_d]^2 
		+\epsilon_d^2 [-4t_c +\omega_c+\omega_d]\Big\}\bigg)\nonumber\\
		&\quad\quad\quad\,\,    -64 [B_z-\omega_c]^2 \bigg(-2 b_x^2 [2t_c -\omega_c] [2t_c
		-\omega_d]^2
		+8 B_z g_c^2 [2t_c -\omega_d]^2+B_z \epsilon_d^2 [2t_c -\omega _c] [4t_c
		-\omega_c-\omega_d] \nonumber\\
		&\quad\quad\quad\quad\quad\quad\quad\quad\quad\quad\quad   -8 g_c^2 \omega_c [2t_c -\omega_d]^2-\omega_c \epsilon_d^2 [2t_c
		-\omega_c] [4t_c -\omega_c-\omega_d]\bigg)\bigg\}\Bigg ]\nonumber\\
		&\quad\quad	\Bigg / \Bigg[32 [\omega_c-B_z] [2t_c -\omega_c]^3
		[2t_c -\omega_d]^2\bigg\{16 [B_z-\omega_c]^2-A^2\bigg\}\Bigg],\\
	\chi=&\frac{A g_c^2 \left[A^3-16 A (B_z-\omega_c)^2+32 b_x^2
		(B_z-\omega_c)\right]}{32 (B_z-\omega_c) (2t_c -\omega_c)^2 \left[16
		(B_z-\omega_c)^2-A^2\right]},\\
	g_{\nu\perp}=
	&\frac{1}{2} g_c\frac{A}{4}\frac{ b_x}{2} \frac{\epsilon_d}{4}	\left(\frac{1}{ \left(\frac{A}{4}+2t_c-\omega_c\right)(2t_c-\omega_c)(B_z-\omega_c)}+\frac{1}{ (2t_c-\omega_d)\left(\frac{A}{4}-B_z+\omega_d\right)\left(\frac{A}{4}-2t_c+\omega_d\right)}\right),\\
	g_{\nu,\mathrm{off}}=&\frac{1}{2}  g_c\frac{A}{4} \frac{b_x}{2} \frac{\epsilon_d}{4}\left(\frac{1}{(2t_c -\omega_c)
		\left(\frac{A}{4}-B_z+\omega_c\right) \left(\frac{A}{4}-2t_c +\omega_c\right)}+\frac{1}{(B_z-\omega_d) (2t_c -\omega_d) \left(\frac{A}{4}+2t_c
		-\omega_d\right)}\right),\\
	g_{\nu\parallel}=&\frac{A g_c \epsilon_d}{256}\left(\frac{A^3+32b_x^2(B_z-\omega_c)-16A(Bz-\omega_c)^2}{(-A^2+16(B_z-\omega_c)^2)(B_z-\omega_c)(2t_c-\omega_c)^2}+\frac{A^3+32b_x^2(B_z-\omega_d)-16A(B_z-\omega_d)^2}{(-A^2+16(B_z-\omega_d)^2)(B_z-\omega_d)(2t_c-\omega_d)^2}\right),
 \label{eq:gnuparallel}\\
	K=&\frac{g_c^4}{(2t_c-\omega_c)^3},\\
		\tilde{b}_x=&\frac{A b_x  \epsilon_d^2 \left(A^2-4 A (2t_c -\omega_d)+32
			(B_z-\omega_d) (2t_c -\omega_d)\right)}{64 (\omega_d-B_z) (2t_c
			-\omega_d) (A-4 B_z+4 \omega_d) \left(16 (2t_c -\omega_d)^2-A^2\right)},\\
		\chi_x=&\frac{A b_x g_c^2  \left(A^2-4 A (2t_c -\omega_c)+32 (B_z-\omega_c)
			(2t_c -\omega_c)\right)}{4 (\omega_c-B_z) (2t_c -\omega_c) (A-4 B_z+4
			\omega_c) \left(16 (2t_c -\omega_c)^2-A^2\right)},\\
		\alpha_9=&\frac{g_c \epsilon _d}{768 \left[ 2 t_c\!-\!\omega
			_c\right]^2} 
		\Biggl[  \frac{6}{\left[ 2t_c\!-\!\omega _d\right]^3} 
		\Bigg\{ \epsilon _d^2\frac{-\!4 t_c^2
			\left[\omega _c\!+\!11 \omega _d\right]\!+\!4 t_c \omega _d \left[\omega _c\!+\!5 \omega _d\right]\!+\!2 \omega _c \omega _d^2\!-\!3
			\omega _c^2 \omega _d\!+\!32 t_c^3\!+\!\omega _c^3\!-\!4 \omega _d^3}{2 t_c\!+\!\omega _c\!-\!2 \omega _d}\nonumber\\
		&\!-\! 64 b_x^2\left[ 2 t_c\!-\!\omega _d\right]
		A^2 \frac{
				-\!B_z \left(\!-\!4 t_c \left[\omega _c\!+\!\omega _d\right]\!+\!8 t_c^2\!+\!\omega
			_c^2\!+\!\omega _d^2\right)\!+\!4 t_c^2 \left[\omega _c\!+\!\omega _d\right]\!-\!8 t_c \omega _c \omega _d\!+\!\omega _c \omega _d
			\left[\omega _c\!+\!\omega _d\right]		
		}
		{ \left(A^2\!-\!16\left[B_z\!-\!\omega _c\right]^2\right) \left(A^2\!-\!16 \left[B_z\!-\!\omega _d\right]^2\right)
		}\nonumber\\
		&\!-\! 1024 b_x^2\left[ 2 t_c\!-\!\omega _d\right] \left[ B_z\!-\!\omega _c\right] \left[ B_z\!-\!\omega _d\right] \frac{
			B_z 
			\left(\!-\!4 t_c \left[\omega _c\!+\!\omega _d\right]\!+\!8 t_c^2\!+\!\omega _c^2\!+\!\omega _d^2
			\right)
			\!-\!4 t_c^2 \left[ \omega _c\!+\!\omega_d\right]\!+\!4 t_c \left[\omega _c^2\!+\!\omega _d^2\right]\!-\!\omega _c^3\!-\!\omega _d^3
		}
		{ \left(A^2\!-\!16\left[B_z\!-\!\omega _c\right]^2\right) \left(A^2\!-\!16 \left[B_z\!-\!\omega _d\right]^2\right)
		}
		\Bigg\}\nonumber\\
		&\!+\!48 A^2 \frac{
			 B_z^2 \left(\!-\!4 t_c \left[\omega _c\!+\!\omega _d\right]\!+\!8 t_c^2\!+\!\omega
			_c^2\!+\!\omega _d^2\right)\!-\!B_z \left(4 t_c^2 \left[3 \omega _c\!+\!\omega _d\right]\!-\!4 t_c \left[\omega _c \omega _d\!+\!2
			\omega _c^2\!+\!\omega _d^2\right]\!+\!\omega _c \omega _d^2\!+\!2 \omega _c^3\!+\!\omega _d^3\right)
		}
		{ \left[B_z\!-\!\omega _d\right] \left[\omega _d\!-\!2 t_c\right]^2 \left(A^2\!-\!16
			\left[B_z\!-\!\omega _c\right]^2\right)
		}\nonumber\\
		&\!+\!\frac{	48A^2
			\left(\omega _c \left\{4 t_c^2\left[\omega _c\!+\!\omega _d\right]\!-\!4 t_c \left[\omega _c^2\!+\!\omega _d^2\right]\!+\!\omega _c^3\!+\!\omega _d^3\right\}\right)
			\!-\!3A^4 \left[\omega _c\!-\!2 t_c\right]^2
		}	
		{ \left[B_z\!-\!\omega _d\right] \left[\omega _d\!-\!2 t_c\right]^2 \left(A^2\!-\!16\left[B_z\!-\!\omega _c\right]^2\right)
		}\nonumber\\
		&\!+\!\frac{3 A^4}{\left[B_z\!-\!\omega _c\right] \left(16 \left[B_z\!-\!\omega_c\right]^2\!-\!A^2\right)} 
		\!+\!24 g_c^2\frac{\!-\!2 t_c \left[7 \omega _c\!+\!9 \omega _d\right]\!-\!\omega _c \omega_d\!+\!32 t_c^2\!+\!4 \omega _c^2\!+\!5 \omega _d^2}{\left[\omega _d\!-\!2 t_c\right]^2(2t_c\!-\!\omega_c)} \!+\!\frac{96 \left[\omega _c\!-\!2t_c\right]^2 \left[\!-\!4 t_c\!+\!\omega _c\!+\!\omega _d\right] }{(2 t_c\!-\!\omega _d)(2t_c\!-\!\omega_c)}
		\Biggr],\\
	\alpha_{11}=&\frac{1}{2} g_c^2  \left(\frac{\epsilon_d}{4}\right)^2\left( -\frac{1}{(2t_c -\omega_c) (\omega_d-2t_c )^2}+\frac{2}{(2t_c -\omega_c)^2 (\omega_d-2t_c )}\right.\nonumber\\
	&\,\,\quad\quad\quad\quad\quad  \left. -\frac{4}{(2t_c -\omega_c)^2 (-2t_c-\omega_c+2 \omega_d)}+\frac{1}{(2t_c -\omega_c)^2 (2t_c -2 \omega_c+\omega_d)}\right),\\
	\alpha_{12}=&\frac{1}{2} g_c^3  \frac{\epsilon_d}{4} \left(\frac{2}{(2t_c -\omega_c) (\omega_d-2t_c )^2}+\frac{1}{(2t_c -\omega_c)^3} +\frac{1}{(2t_c -\omega_c)^2 (2t_c -2 \omega_c+\omega_d)}\right).
	\label{eq:lastcoefficient}
\end{align}
\endgroup
\end{widetext}

\section{Nuclear spin decoherence rate \label{app:nuclear_spin_decoherence_rate}}

In order to numerically estimate the nuclear spin decoherence, or, more precisely, the decoherence rate of the nuclear spin states hybridized with excited charge ($+$) and  electron spin ($\uparrow$) states, we start with the Hamiltonian for the driven QDD system in the rotating reference frame:
\begin{align}
	H^{\mathrm{ R}}_{\mathrm{sys}}=&U_{\mathrm{R,sys}}(t)\left[H_{\mathrm{QDD}}+H_{\mathrm{d}}(t)\right]U^{\dagger}_{\mathrm{R,sys}}(t)\nonumber\\
	&+i \dot{U}_{\mathrm{R,sys}}(t)U^{\dagger}_{\mathrm{ R,sys}}(t)\nonumber\\
	=&\frac{2 t_c-\omega_d}{2}\tau_z+\frac{B_z-\omega_d}{2} \sigma_z+\frac{A}{8}\sigma_z\nu_{z}\nonumber\\
	&-\frac{b_x}{2}\left(\sigma_+ \tau_- +\sigma_- \tau_+\right)\nonumber\\
	&+\frac{A}{4}\left(\sigma_+ \nu_- \tau_- +\sigma_- \nu_+\tau_+\right)-\frac{\epsilon_d}{4}\left(\tau_+ +\tau_-\right).
	\label{eq:HRsys}
\end{align}
Appropriately tuning the parameters of the Hamiltonian results in a situation where the two lowest eigenstates, $\Psi_{\mathrm{sys},1}$ and $\Psi_{\mathrm{sys},2}$, are predominantly the states with anti-aligned nuclear spin while the excited charge ($+$) and  electron spin ($\uparrow$) states contribute only weakly. This is exactly the regime considered in Sec.~\ref{sec:nuclear spin-photon coupling}. For the investigation of the strong coupling condition for nuclear spin photon coupling, the decoherence rate between these two states has to be checked. 
Therefore, in the first step, the unitary operator $U_{\mathrm{sys}}$ that diagonalizes $	H^{\mathrm{ R}}_{\mathrm{sys}}$ is determined and we find the diagonalized form of $H^{\mathrm{ R}}_{\mathrm{sys}}$,
\begin{align}
	H^{\mathrm{R}}_{\mathrm{sys,diag}}=U_{\mathrm{sys}}	H^{\mathrm{ R}}_{\mathrm{sys}} U_{\mathrm{sys}}^{\dagger}. 
\end{align}
Equipped with the diagonalized Hamiltonian, the nuclear spin decoherence dynamics arising due to charge decay and dephasing, as well as electron spin decay and dephasing as a result of the small contribution of the excited orbital and spin states to $\Psi_{\mathrm{sys},1}$ and $\Psi_{\mathrm{sys},2}$ can be inferred from a phenomenological approach based on a Lindblad master equation:
 
\textcolor{Black}{Given a charge qubit with energy splitting $E_{\tau}$ interacting with its environment such that charge decay and charge dephasing occur, these effects can be phenomenologically reproduced by the Lindblad master equation,}
  \begin{align}
 	\dot{\rho}^{\tau}(t)=-i\left[\frac{E_{\tau}}{2}\tau_z,\rho^{\tau}(t)\right]+\gamma^{\tau}\mathcal{D}\left[	\tau_{-}\right]\rho^{\tau}(t)\nonumber\\
 	+\frac{\gamma_{\phi}^{\tau}}{2}\mathcal{D}\left[	\tau_z\right]\rho^{\tau}(t),
 \end{align}
 with the dissipator superoperator ${D[L]\rho(t)}=L\rho(t)L^{\dagger}-\tfrac{1}{2}\left(\rho(t)L^{\dagger}L+L^{\dagger}L\rho(t)\right)$,  the charge decay rate $\gamma^{\tau}$ and the charge dephasing rate $\gamma_{\phi}^{\tau}$. 
 \textcolor{Black}{The solution for this equation for a general initial density matrix,
\begin{align}
    \rho^{\tau}(0)=\begin{pmatrix}
        \rho^{\tau}_{--}(0)& \rho^{\tau}_{-+}(0)\\
         \rho^{\tau}_{+-}(0) & \rho^{\tau}_{++}(0)
    \end{pmatrix},
    \end{align}
    \textcolor{Black}{expressed with respect to the basis of charge qubit states $\{\ket{-},\ket{+}\}$ reads}
 \begin{widetext}
      \begin{align}
     \rho^{\tau}(t)=\begin{pmatrix}
          \rho^{\tau}_{--}(0)+\rho^{\tau}_{++}(0)\left(1- e^{-\gamma^{\tau}t}\right) & \rho^{\tau}_{-+}(0)e^{iE_{\tau}t}e^{-\left(\gamma_{\phi}^{\tau}+\frac{\gamma^{\tau}}{2}\right)t} \\
         \rho^{\tau}_{+-}(0)e^{-iE_{\tau}t}e^{-\left(\gamma_{\phi}^{\tau}+\frac{\gamma^{\tau}}{2}\right)t} & \rho^{\tau}_{++}(0) e^{-\gamma^{\tau}t}
     \end{pmatrix}. 
 \end{align}
 \end{widetext}
 }
 \textcolor{Black}{The population of the excited charge state $\rho^{\tau}_{++}(t)$ decays exponentially with the corresponding decay time $T_1^{\tau}=2\pi/\gamma^{\tau}$. Similarly, the decay of the coherences, $\rho^{\tau}_{-+}(t)$ and $\rho^{\tau}_{+-}(t)$, defines the decoherence time ${T_2^{\tau}=2\pi/\left(\gamma_{\phi}^{\tau}+\tfrac{\gamma^{\tau}}{2}\right)}$. Closely inspecting the expressions for  $T_1^{\tau}$ and  $T_2^{\tau}$, one observes ${T_2^{\tau}\leq 2 \left(2\pi/\gamma^{\tau}\right)=2T_1^{\tau}}$, i.e. $T_2^{\tau}$ is limited by $T_1^{\tau}$. Within the scope of this work we use reported values for the decay $T_1^{\tau}$ and decoherence times $T_2^{\tau}$. These two values are phenomenologically reproduced by choosing the decay and decoherence rates as $\gamma^{\tau}=2\pi/T_1^{\tau}$ and ${\gamma_{\phi}^{\tau}=2\pi\left(1/T_2^{\tau}-1/2T_1^{\tau}\right)}$, respectively. In the limiting case, $T_2^{\tau}=2T_1^{\tau}$, one has ${\gamma_{\phi}^{\tau}=0}$}.
 
 Obviously, the same reasoning can be applied to an electron spin qubit with energy splitting $E_{\sigma}$, electron spin decay time $T^{\sigma}_1$ \textcolor{Black}{($\gamma^{\sigma}=2\pi/T_1^{\sigma}$)} and electron spin decoherence time  $T^{\sigma}_2$ \textcolor{Black}{(${\gamma_{\phi}^{\sigma}=2\pi\left(1/T_2^{\sigma}-1/2T_1^{\sigma}\right)}$)}. Therefore, we argue that the Lindblad master equation,
\begin{align}
	\dot{\rho}_{\mathrm{sys}}(t)=&-i\left[H_{\mathrm{QDD}}+H_{\mathrm{d}}(t),\rho_{\mathrm{sys}}(t)\right]\nonumber\\
	&+\gamma^{\tau}\mathcal{D}\left[	\tau_{-}\right]\rho_{\mathrm{sys}}(t)+\frac{\gamma_\phi^{\tau}}{2}\mathcal{D}\left[\tau_{z} \right]\rho_{\mathrm{sys}}(t)\nonumber\\
	&+\gamma^{\sigma}\mathcal{D}\left[\sigma_{-}\right]\rho_{\mathrm{sys}}(t)+\frac{\gamma_\phi^{\sigma}}{2}\mathcal{D}\left[	\sigma_{z} \right]\rho_{\mathrm{sys}}(t),
\label{eq:rhodotsys}
\end{align}
phenomenologically captures the decoherence dynamics due to charge decay and dephasing as well as electron spin decay and dephasing of the driven QDD system with corresponding density matrix $\rho_{\mathrm{sys}}$. The above master equation does not account for intrinsic nuclear spin decay and dephasing because the reported nuclear spin decay and decoherence times \textcolor{Black}{of ionized $^{31}$P donors in isotopically purified Si} \textcolor{Black}{exceed tens of seconds} \cite{steger2012,pla2013,saeedi2013,muhonen2014} are orders of magnitude larger than those for the charge and electron spin degree of freedom. In order to model the decoherence dynamics in the rotating reference frame defined by $U_{\mathrm{ R,sys}}(t)$ \eqref{eq:URsys}, we multiply both sides of  Eq. \eqref{eq:rhodotsys} with $U_{\mathrm{ R,sys}}(t)$ from the left and $U^{\dagger}_{\mathrm{ R,sys}}(t)$ from the right. In this way, the left hand side of  \eqref{eq:rhodotsys}  transforms as
\begin{align}
	U_{\mathrm{ R,sys}}(t)&\dot{\rho}_{\mathrm{sys}}(t)U^{\dagger}_{\mathrm{R,sys}}(t)=\tfrac{\mathrm{d}}{\mathrm{d}t}\left(U_{\mathrm{ R,sys}}(t)\rho_{\mathrm{sys}}(t)U^{\dagger}_{\mathrm{R,sys}}(t)\right)\nonumber\\
	&-\dot{U}_{\mathrm{ R,sys}}(t)\rho_{\mathrm{sys}}(t)U^{\dagger}_{\mathrm{R,sys}}(t)\nonumber\\
	&-U_{\mathrm{ R,sys}}(t)\rho_{\mathrm{sys}}(t)\dot{U}^{\dagger}_{\mathrm{R,sys}}(t)\nonumber\\
	=&\tfrac{\mathrm{d}}{\mathrm{d}t}\left(U_{\mathrm{ R,sys}}(t)\rho_{\mathrm{sys}}(t)U^{\dagger}_{\mathrm{R,sys}}(t)\right)\nonumber\\
	&+	i \left[-\omega_d\left(\frac{ \tau_z}{2}+\frac{\sigma_z}{2}\right),U_{\mathrm{ R,sys}}(t)\rho_{\mathrm{sys}}(t)U^{\dagger}_{\mathrm{R,sys}}(t)\right]\nonumber\\
	=&\tfrac{\mathrm{d}}{\mathrm{d}t}\left(\rho^{\mathrm{ R}}_{\mathrm{sys}}(t)\right)\nonumber\\
	&+	i \left[i \dot{U}_{\mathrm{R,sys}}(t)U^{\dagger}_{\mathrm{ R,sys}}(t),\rho^{\mathrm{ R,sys}}_{\mathrm{sys}}(t)\right],
	\label{eq:rhorotdot}
\end{align}
where we made use of the explicit form of $U_{\mathrm{ R,sys}}(t)$ \eqref{eq:URsys} in the last two steps and introduced the density matrix in the rotating reference frame ${\rho^{\mathrm{R}}_{\mathrm{sys}}(t)=U_{\mathrm{ R,sys}}(t)\rho_{\mathrm{sys}}(t)U^{\dagger}_{\mathrm{R,sys}}(t)}$.

It is straightforward to verify that the orbital operators obey the transformation rules
\begin{align}
	\tau_z &\rightarrow \tau_z,\nonumber\\
	\tau_+ &\rightarrow  e^{i \omega_d t}\tau_+\nonumber,\\
	\tau_- &\rightarrow  e^{-i \omega_d t}\tau_-\nonumber,\\
\end{align}
and, that the analogous electron spin operators behave similarly. Keeping this in mind, the transformation of the charge decay and dephasing term to the rotating reference frame yields 
\begin{align}
	U_{\mathrm{R}}(t)&\left\{ \mathcal{D}\left[\tau_z\right]\rho_{\mathrm{sys}}(t)\right\}U^{\dagger}_{\mathrm{R}}(t)=\nonumber\\
	&U_{\mathrm{R}}(t)\tau_zU^{\dagger}_{\mathrm{R}}(t)U_{\mathrm{R}}(t)\rho_{\mathrm{sys}}(t)U^{\dagger}_{\mathrm{R}}(t)U_{\mathrm{R}}(t)\tau_zU^{\dagger}_{\mathrm{R}}(t)\nonumber\\
	&-\tfrac{1}{2}\left(U_{\mathrm{R}}(t)\rho_{\mathrm{sys}}(t)U^{\dagger}_{\mathrm{R}}(t)U_{\mathrm{R}}(t)\tau_zU^{\dagger}_{\mathrm{R}}(t)U_{\mathrm{R}}(t)\tau_zU^{\dagger}_{\mathrm{R}}(t)\right.\nonumber\\
	&\left.+U_{\mathrm{R}}(t)\tau_zU^{\dagger}_{\mathrm{R}}(t)U_{\mathrm{R}}(t)\tau_zU^{\dagger}_{\mathrm{R}}(t)U_{\mathrm{R}}(t)\rho_{\mathrm{sys}}(t)U^{\dagger}_{\mathrm{R}}(t)\right)\nonumber\\
	&=\tau_z\rho^{\mathrm{ R}}_{\mathrm{sys}}(t)\tau_z-\tfrac{1}{2}\left(\rho^{\mathrm{ R}}_{\mathrm{sys}}(t)\tau_z\tau_z+\tau_z\tau_z\rho^{\mathrm{ R}}_{\mathrm{sys}}(t)\right)\nonumber\\
	&=\mathcal{D}\left[\tau_z\right]\rho^{\mathrm{ R}}_{\mathrm{sys}}(t), 
	\label{eq:Dtauz}
\end{align}
and
\begin{align}
	U&_{\mathrm{R}}(t)\left\{ \mathcal{D}\left[\tau_-\right]\rho_{\mathrm{sys}}(t)\right\}U^{\dagger}_{\mathrm{R}}(t)=\nonumber\\
	&U_{\mathrm{R}}(t)\tau_-U^{\dagger}_{\mathrm{R}}(t)U_{\mathrm{R}}(t)\rho_{\mathrm{sys}}(t)U^{\dagger}_{\mathrm{R}}(t)U_{\mathrm{R}}(t)\tau_+U^{\dagger}_{\mathrm{R}}(t)\nonumber\\
	&-\tfrac{1}{2}\left(U_{\mathrm{R}}(t)\rho_{\mathrm{sys}}(t)U^{\dagger}_{\mathrm{R}}(t)U_{\mathrm{R}}(t)\tau_+U^{\dagger}_{\mathrm{R}}(t)U_{\mathrm{R}}(t)\tau_-U^{\dagger}_{\mathrm{R}}(t)\right.\nonumber\\
	&\left.+U_{\mathrm{R}}(t)\tau_+U^{\dagger}_{\mathrm{R}}(t)U_{\mathrm{R}}(t)\tau_-U^{\dagger}_{\mathrm{R}}(t)U_{\mathrm{R}}(t)\rho_{\mathrm{sys}}(t)U^{\dagger}_{\mathrm{R}}(t)\right)\nonumber\\
	&=e^{-i\omega_dt}\tau_-\rho^{\mathrm{ R}}_{\mathrm{sys}}(t)e^{i\omega_d t}\tau_+\nonumber\\
	&-\tfrac{1}{2}\left(\rho^{\mathrm{ R}}_{\mathrm{sys}}(t)e^{i\omega_d t}\tau_+e^{-i\omega_d t}\tau_-\right. \nonumber\\
	&\quad\quad\left.+e^{i\omega_d t}\tau_+e^{-i\omega_d t}\tau_-\rho^{\mathrm{ R}}_{\mathrm{sys}}(t)\right)\nonumber\\
	&=\mathcal{D}\left[\tau_-\right]\rho^{\mathrm{ R}}_{\mathrm{sys}}(t),
	\label{eq:Dtauminus}
\end{align}
respectively. Again, the analogous electron spin terms behave similarly.  Finally, with the help of  \eqref{eq:rhorotdot}, \eqref{eq:Dtauz}, \eqref{eq:Dtauminus} and \eqref{eq:HRsys}, the master equation in the rotating reference frame reads
\begin{align}
	\dot{\rho}^{\mathrm{ R}}_{\mathrm{sys}}(t)=&-i\left[H^{\mathrm{R}}_{\mathrm{sys}},\rho^{\mathrm{ R}}_{\mathrm{sys}}(t)\right]\nonumber\\
	&+\gamma^{\tau}\mathcal{D}\left[\tau_{-} \right]\rho(t)+\frac{\gamma_\phi^{\tau}}{2}\mathcal{D}\left[\tau_{z}\right]\rho^{\mathrm{ R}}_{\mathrm{sys}}(t)\nonumber\\
	&+\gamma^{\sigma}\mathcal{D}\left[\sigma_{-} \right]\rho^{\mathrm{ R}}_{\mathrm{sys}}(t)+\frac{\gamma_\phi^{\sigma}}{2}\mathcal{D}\left[	\sigma_{z}\right]\rho^{\mathrm{ R}}_{\mathrm{sys}}(t). 
	\label{eq:rhodotRsys}
\end{align}
This master equation can be further transformed to the basis that diagonalizes $H^{\mathrm{ R}}_{\mathrm{sys}}$:
\begin{align}
	\dot{\rho}^\mathrm{R}_{\mathrm{sys,diag}}(t)=&-i\left[H^{\mathrm{R}}_{\mathrm{sys,diag}},\rho^{\mathrm{ R}}_{\mathrm{sys,diag}}(t)\right]\nonumber\\
	&+\gamma^{\tau}\mathcal{D}\left[U_{\mathrm{sys}}	\tau_{-} U_{\mathrm{sys}}^{\dagger}\right]\rho^{\mathrm{ R}}_{\mathrm{sys,diag}}(t)\nonumber\\
	&+\frac{\gamma_\phi^{\tau}}{2}\mathcal{D}\left[U_{\mathrm{sys}}\tau_{z} U_{\mathrm{sys}}^{\dagger}\right]\rho^{\mathrm{ R}}_{\mathrm{sys,diag}}(t)\nonumber\\
	&+\gamma^{\sigma}\mathcal{D}\left[U_{\mathrm{sys}}	\sigma_{-} U_{\mathrm{sys}}^{\dagger}\right]\rho^{\mathrm{R}}_{\mathrm{sys,diag}}(t)\nonumber\\
	&+\frac{\gamma_\phi^{\sigma}}{2}\mathcal{D}\left[U_{\mathrm{sys}}	\sigma_{z} U_{\mathrm{sys}}^{\dagger}\right]\rho^{\mathrm{R}}_{\mathrm{sys,diag}}(t),
\end{align}
with $	{\rho^\mathrm{R}_{\mathrm{sys,diag}}(t)=U_{\mathrm{sys}}\rho^{\mathrm{ R}}_{\mathrm{sys}}(t)U_{\mathrm{sys}}^{\dagger}}$. With the help of the above master equation the decoherent time evolution of the system initialized in a coherent superposition of the two lowest eigenstates ${\ket{\Psi_{\mathrm{sys}}(0)}=\tfrac{1}{\sqrt{2}}\left(\ket{\Psi_{\mathrm{sys},1}}+\ket{\Psi_{\mathrm{sys},2}}\right)}$, i.e. 
${\rho^\mathrm{R}_{\mathrm{sys,diag}}(0)=\ket{\Psi_{\mathrm{sys}}(0)}\bra{\Psi_{\mathrm{sys}}(0)}}$, can be numerically determined. Then the decoherence rate $\gamma^{\nu}_{\phi}$ between the states $\Psi_{\mathrm{sys},1}$ and $\Psi_{\mathrm{sys},2}$ is encrypted in the time evolution of $f(t)=\Tr\left[\ket{\Psi_{\mathrm{sys},1}}\bra{\Psi_{\mathrm{sys},2}}\rho^\mathrm{R}_{\mathrm{sys,diag}}(t)\right]$. A fit of $|f(t)|$ with the fit function $\tfrac{1}{2}\exp\left(-\frac{\gamma^{\nu}_{\phi}}{2\pi} t\right)$ reveals the searched decoherence rate. For the calculation of the decoherence rates presented in Fig.~\ref{fig:nuclear_spin_photon_coupling}, state-of-the-art values reported for the decay and decoherence times of charge and electron spin qubits realized in $^{28}$Si/SiGe heterostructures are assumed:
\begin{align*}
	\begin{array}{lclcl}
		2\pi/\gamma^{\tau}&=&T_1^{\tau}&=&45\,\upmu\mathrm{s} \,\text{\cite{wang2013a}},\\
		2\pi/\gamma_{\phi}^{\tau}&=&T_2^{\tau,*}&=&0.06\,\upmu\mathrm{s}\, \text{\cite{mi2017}},\\
		2\pi/\gamma^{\sigma} &=&T_1^{\sigma}&=&1\,\mathrm{s}\, \text{\cite{hollmann2020}},\\
		2\pi/\gamma_{\phi}^{\sigma}&=&T_2^{\sigma}&=&3\,\mathrm{ms}\, \text{\cite{yoneda2018}}.
	\end{array}
\end{align*}

\section{Effective Hamiltonian coupled nuclear spin qubits \label{app:coupled_nuclear_spins}}

In order to derive an effective Hamiltonian that describes the interaction of the nuclear spins of two similar driven QDD systems with a microwave resonator mode, \textcolor{Black}{we} follow the procedure outlined in the Sections~\ref{sec:model} and \ref{sec:nuclear spin-photon coupling} as well as Appendix \ref{app:eff_nuclear_spin_photon_coupling}. The system Hamiltonian $\hat{H}(t)$~\eqref{eq:Hhat} is transformed to the rotating reference frame set by $\hat{U}_{\mathrm{ R}}(t)$~\eqref{eq:URhat}:
\begin{align}
	\hat{H}^{\mathrm{R}}=&	\hat{U}_{\mathrm{R}}(t)	\hat{H}(t)\hat{U}^{\dagger}_{\mathrm{R}}(t)+i \dot{\hat{U}}_{\mathrm{R}}(t)\hat{U}^{\dagger}_{\mathrm{R}}(t)\nonumber\\
	=&\hat{H}^{\mathrm{R}}_{0}+\hat{V}^{\mathrm{R}},
	\label{eq:HRhat}
\end{align}
whereby the result can be partitioned in a diagonal part $\hat{H}^{\mathrm{ R}}_{0}$ and an off-diagonal part $\hat{V}^{\mathrm{R}}$, 
\begin{align}
	\hat{H}^{\mathrm{ R}}_{0}=\sum_{i=1}^{2}&\left[\frac{2 t_c-\omega_d}{2}\tau_z^{(i)}+\frac{B_z-\omega_d}{2} \sigma_z^{(i)}+\frac{A}{8}\sigma_z^{(i)}\nu_{z}^{(i)}\right]\nonumber\\
	&+\left(\omega_c-\omega_d\right) a^{\dagger}a,\\
	\hat{V}^{\mathrm{R}}=&\sum_{i=1}^{2}\left[-\frac{b_x}{2}\left(\sigma^{(i)}_+ \tau^{(i)}_- +\sigma^{(i)}_- \tau^{(i)}_+\right)\right.\nonumber\\
	&\quad\quad\left.-g_c\left(\tau^{(i)}_+a+\tau^{(i)}_-a^{\dagger}\right)\right.\nonumber\\
	&\quad\quad\left.+\frac{A}{4}\left(\sigma^{(i)}_+ \nu^{(i)}_- \tau^{(i)}_- +\sigma^{(i)}_- \nu^{(i)}_+\tau^{(i)}_+\right)\right.\nonumber\\
	&\quad\quad\left.-\frac{\epsilon_d}{4}\left(\tau^{(i)}_+ +\tau^{(i)}_-\right)\right].
	\label{eq:VRhat}
\end{align}
As the next step, we apply a Schrieffer-Wolff transformation to find an effective Hamiltonian describing the dynamics of the nuclear spins interacting with the resonator mode, i.e. the subspace defined by the projection operator $\hat{P}_0$~\eqref{eq:P0hat}. Considering terms up to fourth order in the perturbation $\hat{V}^{\mathrm{R}}$ we find  
\begin{align}
	\hat{H}^{\mathrm{R}}_{\nu-\mathrm{ph}}
	=&\sum_{i=1}^{2}\left[-\frac{E_{\nu}}{2}\nu_z^{(i)}+\chi\nu_z^{(i)} a^{\dagger}a\right.\nonumber\\
	&\left.\quad\quad+g_{\nu\perp}\left(\nu_-^{(i)} a+\nu_+^{(i)} a^{\dagger}\right)\right.\nonumber\\
	&\quad\quad+g_{\nu,\mathrm{off}}\left(\nu_+^{(i)} a+\nu_-^{(i)} a^{\dagger}\right)+g_{\nu\parallel}\nu_z^{(i))}\left(a+a^{\dagger}\right)\nonumber\\
	&\quad\quad\left.+\left(\tilde{b}_x  +\chi_x  a^{\dagger}a\right)\nu_x^{(i)}\right]\nonumber\\
	&+\bar{\omega}_c a^{\dagger}a+\bar{K}\left(a^{\dagger}a\right)^2+\bar{\alpha}_9 \left(a+a^{\dagger}\right)\nonumber\\
	&+\bar{\alpha}_{11} \left(a^2+(a^{\dagger})^2\right)+\bar{\alpha}_{12}\left(a^{\dagger}a^2+\left(a^{\dagger}\right)^2a\right).
\end{align}
This effective Hamiltonian has a similar structure as the one derived for the single QDD system interacting with the microwave resonator $H^{\mathrm{ R}}_{\nu-\mathrm{ph}}$~\eqref{eq:Heffappnedix}. The various coefficients agree with those given in the Eqs.~\eqref{eq:firstcoefficient}~-~\eqref{eq:lastcoefficient} unless marked with a bar. In the latter case the coefficients are supplemented by additional perturbative corrections originating from processes that involve both QDD systems:
\begingroup
\allowdisplaybreaks
\begin{align}
	\bar{\omega}_c=&2\tilde{\omega}_c-\left(\omega_c-\omega_d\right)+\frac{2g_c^4}{\left(2t_c-\omega_c\right)^3}\nonumber\\
	&+\frac{g_c^2\epsilon_d^2}{6}\left(\frac{1}{\left(2t_c-\omega_c\right)\left(2t_c-\omega_d\right)^2}\right.\nonumber\\
	&\quad\quad\quad\quad\left.+\frac{1}{\left(2t_c-\omega_c\right)^2\left(2t_c-\omega_d\right)}\right),\\
	\bar{K}=&2K+\frac{4g_c^4}{3\left(2t_c-\omega_c\right)^3},
	\label{eq:baralpha4}\\
	\bar{\alpha}_9=&2\alpha_9+\frac{17g_c^3\epsilon_d}{48\left(2t_c-\omega_c\right)^3}+\frac{g_c\epsilon_d^3}{96 \left(2t_c-\omega_d\right)^3}\nonumber\\
	&+\frac{8g_c^3\epsilon_d+g_c\epsilon_d^3}{32\left(2t_c-\omega_c\right)\left(2t_c-\omega_d\right)^2}\nonumber\\
	&+\frac{11g_c^3\epsilon_d}{48\left(2t_c-\omega_c\right)^2\left(2t_c-\omega_d\right)},\\
	\bar{\alpha}_{11}=&2\alpha_{11}+\frac{g_c^2\epsilon_d^2}{24}\left(\frac{1}{\left(2t_c-\omega_c\right)\left(2t_c-\omega_d\right)^2}\right.\nonumber\\
	&\quad\quad\quad\quad\quad\quad\left.+\frac{1}{\left(2t_c-\omega_c\right)^2\left(2t_c-\omega_d\right)}\right),\\
	\bar{\alpha}_{12}=&2\alpha_{12}+\frac{g_c^3\epsilon_d}{6}\left(\frac{1}{\left(2t_c-\omega_c\right)^3}\right.\nonumber\\
	&\quad\quad\quad\quad\quad\quad\left.+\frac{3}{\left(2t_c-\omega_c\right)^2\left(2t_c-\omega_d\right)}\right).
\end{align}
\endgroup

\subsection{Effective nuclear spin interaction}
In order to perform a Schrieffer-Wolff transformation that yields an effective Hamiltonian for the subspace defined by the projection operator $\hat{P}_0^{\nu-\nu}$, we first have to  divide $\hat{H}^{\mathrm{ R}}_{\nu-\mathrm{ph}}$ in a diagonal part ($\hat{H}^{\mathrm{ R}}_{\nu-\mathrm{ph},0}$), the block off-diagonal perturbation ($	\hat{V}^{\mathrm{ R}}_{\nu-\mathrm{ph},\mathrm{od}}$) and the block diagonal perturbation ($	\hat{V}^{\mathrm{ R}}_{\nu-\mathrm{ph},\mathrm{d}}$) according to the description of the formalism in Appendix \ref{app:eff_nuclear_spin_photon_coupling}. $\hat{H}_0$ is given by
\begin{align}
\hat{H}^{\mathrm{ R}}_{\nu-\mathrm{ph},0}&=\sum_{i=1}^{2}\left[-\frac{E_{\nu}}{2}\nu_z^{(i)}+\chi\nu_z^{(i)} \left\{\sum_{n=1}^{\infty}n \ket{n}\bra{n}\right\}\right]\nonumber\\
	&+\bar{\omega}_c  \left\{\sum_{n=1}^{\infty}n \ket{n}\bra{n}\right\}+\bar{K} \left\{\sum_{n=1}^{\infty}n^2 \ket{n}\bra{n}\right\},
	\label{eq:hatH0}
\end{align}
with the resonator mode photon number states ${\ket{n},\,\,n=0,1,2,...}$. Using the projection operators $\hat{P}_0^{\nu-\nu}$ and ${\hat{Q}_0^{\nu-\nu}=1-\hat{P}_0^{\nu-\nu}}$ we find
\begin{widetext}
\begin{align}
\hat{V}^{\mathrm{ R}}_{\nu-\mathrm{ph},\mathrm{d}}=&\hat{P}_0^{\nu-\nu} (\hat{H}_{\mathrm{eff}}-\hat{H}_0)\hat{P}_0^{\nu-\nu}+\hat{Q}_0^{\nu-\nu} (\hat{H}_{\mathrm{eff}}-\hat{H}_0)\hat{Q}_0^{\nu-\nu} \nonumber\\
	=&\sum_{i=1}^{2}\left[g_{\nu\perp}\left(\nu_-^{(i)} 	\left\{\sum_{n=2}^{\infty}\sqrt{n}\ket{n-1}\bra{n}\right\}+\nu_+^{(i)} \left\{\sum_{n=2}^{\infty}\sqrt{n}\ket{n}\bra{n-1}\right\}\right)\right.\nonumber\\
	&\quad\quad+g_{\nu,\mathrm{off}}\left(\nu_+^{(i)} \left\{\sum_{n=2}^{\infty}\sqrt{n}\ket{n-1}\bra{n}\right\}+\nu_-^{(i)} 	\left\{\sum_{n=2}^{\infty}\sqrt{n}\ket{n}\bra{n-1}\right\}\right) \nonumber\\
	&\quad\quad+g_{\nu\parallel}\nu_z^{(i)}\left\{\sum_{n=2}^{\infty}\sqrt{n}\left(\ket{n-1}\bra{n}+\ket{n}\bra{n-1}\right)\right\} 	\nonumber\\
	&\quad\quad\left.+\left(\tilde{b}_x  +\chi_x  \left\{\sum_{n=1}^{\infty}n \ket{n}\bra{n} \right\}\right)\nu_x^{(i)}\right]\nonumber\\
	&+\bar{\alpha}_9 \left\{\sum_{n=2}^{\infty}\sqrt{n}\left(\ket{n-1}\bra{n}+\ket{n}\bra{n-1}\right)\right\}\nonumber\\
	&+\bar{\alpha}_{11} \left\{\sum_{n=3}^{\infty}\sqrt{n (n-1)}\left(\ket{n-2}\bra{n}+\ket{n}\bra{n-2}\right)\right\}\nonumber\\
	&+\bar{\alpha}_{12}	\left\{\sum_{n=2}^{\infty}(n-1)\sqrt{n}\left(\ket{n-1}\bra{n}+\ket{n}\bra{n-1}\right)\right\}, 
\end{align}
\end{widetext}
and 
\begin{align}
	\hat{V}^{\mathrm{ R}}_{\nu-\mathrm{ph},\mathrm{od}}=&\hat{P}_0^{\nu-\nu} (\hat{H}_{\mathrm{eff}}-\hat{H}_0)\hat{Q}_0^{\nu-\nu}\nonumber\\
	&+\hat{Q}_0^{\nu-\nu} (\hat{H}_{\mathrm{eff}}-\hat{H}_0)\hat{P}_0^{\nu-\nu} \nonumber\\
	=&\sum_{i=1}^{2}\left[g_{\nu\perp}\left(\nu_-^{(i)} \ket{0}\bra{1}+\nu_+^{(i)}  \ket{1}\bra{0}\right)\right.\nonumber\\
	&\left.\quad\quad+g_{\nu,\mathrm{off}}\left(\nu_+^{(i)}  \ket{0}\bra{1}+\nu_-^{(i)}  \ket{1}\bra{0}\right)\right.\nonumber\\
	&\left. \quad\quad+g_{\nu\parallel}\nu_z^{(i))}\left(\ket{0}\bra{1}+\ket{1}\bra{0}\right)\right]\nonumber\\
	&+\bar{\alpha}_9 \left(\ket{0}\bra{1}+\ket{1}\bra{0}\right)\nonumber\\
	&+\bar{\alpha}_{11} \sqrt{2} \left(\ket{0}\bra{2}+\ket{2}\bra{0}\right).
	\label{eq:hatVod}
\end{align}
At this point we introduce a photon number cutoff $n_{\mathrm{max}}$ by discarding all summands with $n>n_{\mathrm{max}}$ in the infinite sums in \eqref{eq:hatH0}-\eqref{eq:hatVod}. This simplifies the determination of the generator of the Schrieffer-Wolff transformation and finally the effective Hamiltonian.
 We note that all the perturbative corrections contributing to the effective Hamiltonian originate from coupling sequences, as the ones shown in the Figs.~\ref{fig:coupling_scheme} and \ref{fig:effectivecouplingmechanismnuclearspins}, with the initial and final state in the subspace set by $\hat{P}_0^{\nu-\nu}$ and at least one intermediate state in the subspace defined by $\hat{Q}_0^{\nu-\nu}$. Therefore any of the coupling sequences is at least second order in $\hat{V}^{\mathrm{ R}}_{\nu-\mathrm{ph},\mathrm{od}}$ and coupling sequences of length $l_s$ are at most $(l_s-2)$-th order in $\hat{V}_{\mathrm{d}}$. The length $l_s$ of such a coupling sequence sets the order in the perturbation ${\hat{V}^{\mathrm{ R}}_{\nu-\mathrm{ph}}=\hat{V}^{\mathrm{ R}}_{\nu-\mathrm{ph},\mathrm{od}}+\hat{V}^{\mathrm{R}}_{\nu-\mathrm{ph},\mathrm{d}}}$ with which the corresponding term contributes to the effective Hamiltonian. In the present case $\hat{V}_{\mathrm{od}}$ couples states in the subspace corresponding to $\hat{P}_0^{\nu-\nu}$ to states with at most two photons in the subspace set by $\hat{Q}_0^{\nu-\nu}$ implying that the first and the last intermediate state in the subspace defined by $\hat{Q}^{\nu-\nu}_0$ in any sequence is at most a two photon state. On the other hand $\hat{V}_d$ contains couplings that either raise or lower the photon number by at most two. For a given length $l_s$ there are sequences with the first and the last intermediate state being a two photon state and the remaining intermediate states all lying in the subspace defined by $\hat{Q}_0$ and, therefore, the remaining $l_s-2$ couplings are all governed by $\hat{V}^{\mathrm{ R}}_{\nu-\mathrm{ph},\mathrm{d}}$. Assuming that $l_s$ is even, $(l_s-2)/2$ of the remaining couplings can sequentially raise the photon number of the intermediate states by two, such that the $(l_s/2)$-th intermediate state is a $2+2 [(ls-2)/2]$ photon state. Then, the remaining $(l_s-2)/2$ couplings have to lower the photon number by two sequentially to reach the last intermediate state, a two photon state. This train of thoughts shows that, in the present case, states with photon numbers larger than $2+2 [(ls-2)/2]$ do not affect the result of the Schrieffer-Wolff transformation up to $l_s$-th order in the perturbation $\hat{V}^{\mathrm{ R}}_{\nu-\mathrm{ph}}$ for even $l_s$. In a similar way, one finds that this holds for states with photon number  $2+2 [(\{ls-1\}-2)/2]$ for odd $l_s$.  Thus, for a Schrieffer Wolff transformation up to $l_s$-th order in the perturbation $\hat{V}$ we can set the photon number cutoff 
 \begin{align}
 	n_{\mathrm{max}}=\left\{
 		\begin{array}{l l}
 		2+2 [(ls-2)/2] & \text{for $l_s$ even}\\
 		2+2 [(\{ls-1\}-2)/2]  & \text{for $l_s$ odd}
 	\end{array}
 	\right.
 \end{align}
without affecting the result obtained for the effective Hamiltonian due to the chosen cutoff. Hence, for the determination of the effective Hamiltonian up to second order in the perturbation $\hat{V}$ it is sufficient to consider two photons and we find
\begin{align}
	\hat{H}^{\mathrm{R}}_{\nu-\nu}=&-\sum_{i=1}^{2}\frac{\bar{E}_{\nu}}{2}\,\nu_z^{(i)}+\delta \bar{E}_{\nu}\,\nu_z^{(1)}\nu_z^{(2)}\nonumber\\
	&+\sum_{i=1}^2\left(\bar{b}_x+\sum_{j\neq i}c_2\nu_z^{(j)} \right)\nu_x^{(i)}\nonumber\\
	&+\zeta_{\nu\nu\perp}\left(\nu_+^{(1)}\nu_-^{(2)}+\nu_-^{(1)}\nu_+^{(2)}\right)\nonumber\\
	&+\zeta_{\nu\nu\parallel}\left(\nu_+^{(1)}\nu_+^{(2)}+\nu_-^{(1)}\nu_-^{(2)}\right),
\end{align}
with
\begingroup
\allowdisplaybreaks
\begin{align}
	\bar{E}_{\nu}=&E_{\nu}-\frac{g_{\nu\perp}^2}{\bar{K}-E_{\nu}+\omega_c}+\frac{g_{\nu,\mathrm{off}}^2}{\bar{K}+E_{\nu}+\omega_c}\nonumber\\
	&-\frac{\bar{\alpha}_{11}^2}{2\left(2\bar{K}-2\chi+\bar{\omega}_c\right)}+\frac{\bar{\alpha}_{11}^2}{2\left(2\bar{K}+2\chi+\bar{\omega}_c\right)}\nonumber\\
	&-\frac{\left(\bar{\alpha}_9-2g_{\nu\parallel}\right)^2}{2\left(\bar{K}-2\chi+\bar{\omega}_c\right)}+\frac{\left(\bar{\alpha}_9+2g_{\nu\parallel}\right)^2}{2\left(\bar{K}+2\chi+\bar{\omega}_c\right)},\\
	\delta\bar{E}_{\nu}=&\frac{\bar{\alpha}_9^2}{2\left(\bar{K}+\bar{\omega}_c\right)}+\frac{\bar{\alpha}_{11}^2}{2\left(2\bar{K}+\bar{\omega}_c\right)}-\frac{\bar{\alpha}_{11}^2}{4\left(2\bar{K}+2\chi+\bar{\omega}_c\right)}\nonumber\\
	&-\frac{\bar{\alpha}_{11}^2}{4\left(2\bar{K}-2\chi+\bar{\omega}_c\right)}\nonumber\\
	&-\frac{g_{\nu\perp}^2}{2\left(\bar{K}-E_{\nu}+\bar{\omega}_c\right)}
	+\frac{g_{\nu\perp}^2}{2\left(\bar{K}-E_{\nu}+\bar{\omega}_c+2\chi\right)}\nonumber\\
	&-\frac{g_{\nu,\mathrm{off}}^2}{2\left(\bar{K}+E_{\nu}+\bar{\omega}_c\right)}+\frac{g_{\nu,\mathrm{off}}^2}{2\left(\bar{K}+E_{\nu}-2\chi+\bar{\omega}_c\right)}\nonumber\\
	&-\frac{\left(\bar{\alpha}_9-2g_{\nu\parallel}\right)^2}{4\left(\bar{K}-2\chi+\bar{\omega}_c\right)}-\frac{\left(\bar{\alpha}_9+2g_{\nu\parallel}\right)^2}{4\left(\bar{K}+2\chi+\bar{\omega}_c\right)}, \label{eq:deltaEbarnu} \\
	\bar{b}_x=&\tilde{b}_x+\frac{\bar{\alpha}_9g_{\nu\perp}}{4\left(-\bar{K}+E_{\nu}-\bar{\omega}_c\right)}+\frac{g_{\nu\perp}\left(\bar{\alpha}_9+2g_{\nu\parallel}\right)}{4\left(-\bar{K}+E_{\nu}-2\chi-\bar{\omega}_c\right)}\nonumber\\
	&+\frac{g_{\nu,\mathrm{off}}\left(\bar{\alpha}_9-2g_{\nu\parallel}\right)}{4\left(-\bar{K}+2\chi-\bar{\omega}_c\right)}+\frac{g_{\nu,\mathrm{off}}\left(2g_{\nu\parallel}-\bar{\alpha}_9\right)}{4\left(\bar{K}+E_{\nu}-2\chi+\bar{\omega}_c\right)}\nonumber\\
	&-\frac{\bar{\alpha}_9\left(g_{\nu\perp}+g_{\nu,\mathrm{off}}\right)}{4\left(\bar{K}+\bar{\omega}_c\right)}-\frac{\bar{\alpha}_9g_{\nu,\mathrm{off}}}{4\left(\bar{K}+E_{\nu}+\bar{\omega}_c\right)}\nonumber\\
	&-\frac{g_{\nu\perp} \left(\bar{\alpha}_9+g_{\nu\parallel}\right)}{4\left(\bar{K}+2\chi+\bar{\omega}_c\right)},\\
	c_2=&-\frac{\bar{\alpha}_9g_{\nu\perp}}{4\left(-\bar{K}+E_{\nu}-\bar{\omega}_c\right)}+\frac{g_{\nu\perp}\left(\bar{\alpha}_9+2g_{\nu\parallel}\right)}{4\left(-\bar{K}+E_{\nu}-2\chi-\bar{\omega}_c\right)}\nonumber\\
	&-\frac{g_{\nu,\mathrm{off}}\left(\bar{\alpha}_9-2g_{\nu\parallel}\right)}{4\left(-\bar{K}+2\chi-\bar{\omega}_c\right)}-\frac{g_{\nu,\mathrm{off}}\left(2g_{\nu\parallel}-\bar{\alpha}_9\right)}{4\left(\bar{K}+E_{\nu}-2\chi+\bar{\omega}_c\right)}\nonumber\\
	&+\frac{\bar{\alpha}_9\left(g_{\nu\perp}-g_{\nu,\mathrm{off}}\right)}{4\left(\bar{K}+\bar{\omega}_c\right)}-\frac{\bar{\alpha}_9g_{\nu,\mathrm{off}}}{4\left(\bar{K}+E_{\nu}+\bar{\omega}_c\right)}\nonumber\\
	&-\frac{g_{\nu\perp} \left(\bar{\alpha}_9+g_{\nu\parallel}\right)}{4\left(\bar{K}+2\chi+\bar{\omega}_c\right)},\\
	\zeta_{\nu\nu\perp}=&\frac{g_{\nu\perp}^2}{E_{\nu}-2\chi-\bar{\omega}_c-\bar{K}}-\frac{g_{\nu,\mathrm{off}}^2}{E_{\nu}-2\chi+\bar{\omega}_c+\bar{K}},\\
	\zeta_{\nu\nu,\mathrm{off}}=&\frac{g_{\nu\perp}g_{\nu,\mathrm{off}}}{E_{\nu}-\bar{\omega}_c-\bar{K}}-\frac{g_{\nu\perp}g_{\nu,\mathrm{off}}}{E_{\nu}+\bar{\omega}_c+\bar{K}}.
\end{align}
\endgroup

\section{Justification of the rotating wave approximation (RWA) \label{app:justification_RWA}}

\begin{figure}
	\centering
 \includegraphics[width=\columnwidth]{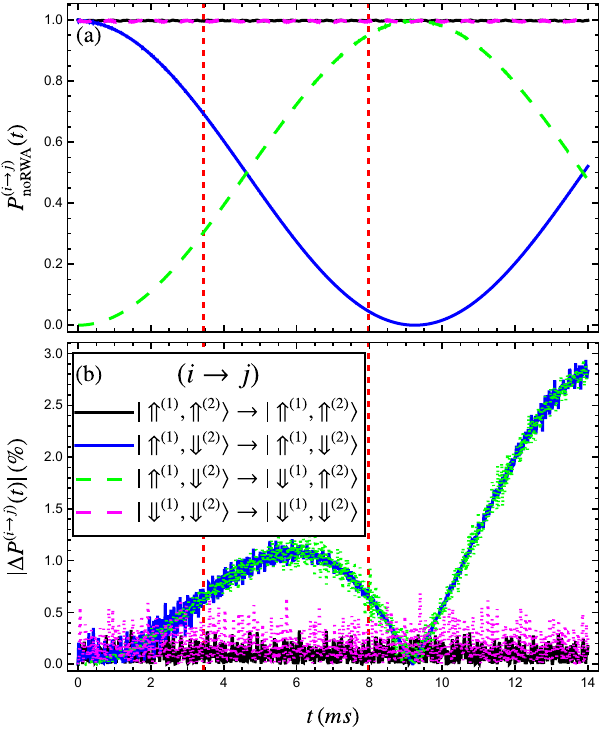}
 \caption{\textcolor{Black}{(a) Transition probability between the various nuclear spin states (as indicated in the legend) without the RWA applied plotted as a function of the evolution time. (b) Difference in the transition probabilities between the time evolution governed by the Hamiltonian with and without the RWA applied. In both plots the first and second vertical dashed red lines indicate the gate times for the $\sqrt{i\mathrm{SWAP}}$-gate and 
 $i\mathrm{SWAP}$-gate found in Appendix~\ref{app:effective decoherence dynamics nuclear spin gate }, respectively. The system parameters are the ones listed in Appendix~\ref{app:effective decoherence dynamics nuclear spin gate }.}}
 \label{fig:RWA_justification}
 \end{figure}

\textcolor{Black}{The small effective coupling between two nuclear spins and the resulting long gate times for the $\sqrt{i\mathrm{SWAP}}$ and the $i\mathrm{SWAP}$ quantum gates imposes the question whether the rotating wave approximation underlying the reasoning in the main text is still justified. }

\textcolor{Black}{First, we note that $\widetilde{H}_{\mathrm{DQD}}$ (Eq.~\eqref{eq:HtildeQDD}), $\widetilde{H}_{\mathrm{int}}$ (Eq.~\eqref{eq:Htildeint}) and $\widetilde{H}_{d}$ (Eq.~\eqref{eq:Htilded}) transformed to the $\ket{\pm}$-basis without applying the RWA read 
\begin{align}
	\breve{H}_{\mathrm{QDD}}=&\frac{1}{2}\left[2t_c\tau_z+B_z \sigma_z +\frac{A}{4}\sigma_z\nu_z-b_x\sigma_x\tau_x\right]\nonumber\\
	&+\frac{A}{8}\sigma_z\nu_z\tau_x+\frac{A}{4}\left(\sigma_+\nu_-+\sigma_-\nu_+\right)\left(1+\tau_x\right),
	\label{eq:HbreveQDD}\\
	\breve{H}_{\mathrm{int}}=&-g_c\left(a+a^{\dagger}\right)\tau_x,\\
	\breve{H}_{d}(t)=&-\frac{\epsilon_d}{2} \cos\left(\omega_d t\right)\tau_x.
	\label{eq:Hbreved}
\end{align}
Hence, two identical driven QDD systems interacting with the same microwave resonator can be modelled by the Hamiltonian
\begin{align}
    \check{H}(t)=\sum_{i=1}^2\left(\breve{H}_{\mathrm{QDD}}^{(i)}+\breve{H}_{\mathrm{int}}^{(i)}+\breve{H}_{d}^{(i)}(t)\right)+H_{\mathrm{cav}}.
\end{align}
In line with the discussion in the main text this Hamiltonian can be transformed to the rotating reference frame defined by $\hat{U}_{\mathrm{R}}(t)$ (Eq.~\eqref{eq:URhat}):
\begin{align}
    \check{H}^{\mathrm{R}}(t)=&\hat{U}_{\mathrm{R}}(t)  \check{H}(t)\hat{U}^{\dagger}_{\mathrm{R}}(t)+i\dot{\hat{U}}_{\mathrm{R}}(t)\hat{U}^{\dagger}_{\mathrm{R}}(t)\nonumber\\
    =&\sum_{i=1}^2\left(\breve{H}_{\mathrm{QDD}}^{\mathrm{R},(i)}(t)+\breve{H}_{\mathrm{int}}^{\mathrm{R},(i)}(t)+\breve{H}_{d}^{\mathrm{R},(i)}(t)\right)+\breve{H}^{\mathrm{R}}_{\mathrm{cav}},
\end{align}
with 
\begingroup
\allowdisplaybreaks
\begin{align}
	\breve{H}_{\mathrm{QDD}}^{\mathrm{R},(i)}(t)=&\frac{1}{2}\Big[\left.\left(2t_c-\omega_d\right)\tau_z^{(i)}+\left(B_z-\omega_d\right)\sigma_z^{(i)} \right.\nonumber\\
& \left.\quad+\frac{A}{4}\sigma_z^{(i)}\nu_z^{(i)}\right. 
 \left.-b_x\left(\sigma_+^{(i)}e^{i\omega_d t}+\sigma_-^{(i)}e^{-i \omega_d t}\right)\right.\nonumber\\
 &\quad\left. \times\left(\tau_+^{(i)}e^{i\omega_d t}+\tau_-^{(i)}e^{-i \omega_d t}\right)\right.\Big]\nonumber\\
	&+\frac{A}{8}\sigma_z^{(i)}\nu_z^{(i)}\left(\tau_+^{(i)}e^{i\omega_d t}+\tau_-^{(i)}e^{-i \omega_d t}\right)\nonumber\\
 &+\frac{A}{4}\left(\sigma_+^{(i)}e^{i\omega_d t}\nu_-^{(i)}+\sigma_-^{(i)}e^{-i\omega_d t}\nu_+^{(i)}\right)\nonumber\\
 &\quad\times\left(1+\tau_+^{(i)}e^{i\omega_d t}+\tau_-^{(i)}e^{-i \omega_d t}\right),
 	\label{eq:HQDDbreveR}\\
	\breve{H}_{\mathrm{int}}^{\mathrm{R},(i)}(t)&=-g_c\left(ae^{-i\omega_d t}+a^{\dagger}e^{i\omega_dt}\right)\nonumber\\
 &\quad\quad\times\left(\tau_+^{(i)}e^{i\omega_d t}+\tau_-^{(i)}e^{-i \omega_d t}\right), \\
	\breve{H}_{d}^{\mathrm{R},(i)}(t)&=-\frac{\epsilon_d}{2} \cos\left(\omega_d t\right)\left(\tau_+^{(i)}e^{i\omega_d t}+\tau_-^{(i)}e^{-i \omega_d t}\right),
	\label{eq:HdbreveR} \\
 \breve{H}^{\mathrm{R}}_{\mathrm{cav}}&=\left(\omega_c-\omega_d\right)a^{\dagger}a.
 \label{eq:HcavR}
\end{align}
\endgroup
The Eqs.~\eqref{eq:HQDDbreveR} to \eqref{eq:HcavR} unveil that $\check{H}^{\mathrm{R}}(t)$ is periodic in time, i.e. $ {\check{H}^{\mathrm{R}}(t)=\check{H}^{\mathrm{R}}(t+T)}$, with period ${T=2\pi/\omega_d}$. We make use of this property by applying Floquet theory \cite{dittrich1998} to numerically calculate the stroboscopic time evolution operator for integer multiples of the period $T$,
\begin{align}
    \check{U}_{t}^{\mathrm{R}}(nT)=\left[\check{U}_{t}^{\mathrm{R}}(T)\right]^n.
\end{align}
Equipped with the stroboscopic time evolution operator, it is straightforward to determine the stroboscopic transition probabilities
\begin{align}
    \textcolor{Black}{P^{(i\rightarrow j)}_{\mathrm{noRWA}}(nT)=\bra{j}\check{U}_{t}^{\mathrm{R}}(nT)\ket{i}},
\end{align}
with $i,j\in\{1,2,3,4\}$. The numbers label the four lowest energy eigenstates of $\breve{H}_{\mathrm{QDD}}^{(1)}+\breve{H}_{\mathrm{QDD}}^{(2)}$ that correspond to the four combinations of nuclear spin states for two QDD systems, i.e. ${\ket{1}\hat{=}\ket{\Uparrow^{(1)},\Uparrow^{(2)}}}$, ${\ket{2}\hat{=}\ket{\Uparrow^{(1)},\Downarrow^{(2)}}}$,
${\ket{3}\hat{=}\ket{\Downarrow^{(1)},\Uparrow^{(2)}}}$ and ${\ket{4}\hat{=}\ket{\Downarrow^{(1)},\Downarrow^{(2)}}}$. It is noteworthy that these states are not affected by the transformation to the rotating reference frame at the considered stroboscopic instances in time because
\begin{align}
    \hat{U}_{\mathrm{R}}(nT)=\mathbb{1}.
\end{align}
Figure~\ref{fig:RWA_justification}a shows that, qualitatively, one observes the expected behaviour  without applying the RWA: there is population transfer between the states with anti-aligned nuclear spins, while the population of the states with parallel nuclear spins remains constant.}  

\textcolor{Black}{For a more quantitative comparison between the time evolution with and without the RWA applied, we also introduce the transition probabilities in case that the RWA is applied
\begin{align}
    \textcolor{Black}{P^{(i\rightarrow j)}_{\mathrm{RWA}}(nT)=\bra{j}e^{-i\hat{H}^{\mathrm{R}}nT}\ket{i}}, 
\end{align}
with $\hat{H}^{\mathrm{R}}$ defined in \eqref{eq:HRhat}.
Here, the states with $i,j\in\{1,2,3,4\}$ are the four lowest energy eigenstates of ${H_{\mathrm{QDD}}^{(1)}+H_{\mathrm{QDD}}^{(2)}}$ and can be identified by the nuclear spin states as above. Finally, we define the transition probability difference between the time evolution with and without the RWA,
\begin{align}
    |\Delta P^{(i\rightarrow j)}(nT)|=|P^{(i\rightarrow j)}_{\mathrm{noRWA}}(nT)-P^{(i\rightarrow j)}_{\mathrm{RWA}}(nT)|.
\end{align}
This quantity is plotted in Fig.~\ref{fig:RWA_justification}b. One observes that the deviation between the two cases does not exceed $\approx 1\%$ for times smaller than the gate time of the $i\mathrm{SWAP}$ gate (to the left of the second dashed vertical red line). Therefore, we conclude that it is justified to apply the RWA in the context of this work.}

\section{Effective decoherence model \label{app:effective decoherence model}}
We aim at numerically determining the average gate fidelity for both the $i\mathrm{SWAP}$ and the $\sqrt{i\mathrm{SWAP}}$ gate in the presence of decoherence effects. Thereby we again choose the phenomenological approach introduced in Appendix~\ref{app:nuclear_spin_decoherence_rate} in order to consider charge decay and dephasing, electron  spin decay and dephasing as well as resonator photon decay. However, solving the Lindblad master equation for the full system consisting of two QDD systems each contributing with a charge, a electron spin and a nuclear spin degree of freedom as well as the microwave resonator mode with photon numbers from 0 up to a cutoff number would require large computational resources due to the large Hilbert space. Therefore, we first consider only a single nuclear spin of a QDD system dispersively interacting with the resonator mode. The master equation for this system can be numerically solved with reasonable computational effort. In the next step we demonstrate that the decoherence dynamics of the system initialized in the nuclear spin subspace can be modelled by a three level model including the two nuclear spin states and a leakage state with high accuracy. 

The effective decoherence model is then applied to the relevant configuration with the nuclear spins of two QDD systems dispersively interacting with the microwave resonator mode. Using this effective decoherence model allows the determination of the average gate fidelity for the $\sqrt{i\mathrm{SWAP}}$ and the $i\mathrm{SWAP}$ quantum gate between the two nuclear spin qubits in the presence of decoherence effects.

\subsection{Effective nuclear spin decoherence model of a single QDD system interacting with a cavity mode  \label{app:eff_decoherence_single_QDD}}

 \begin{figure}[!t]
 \centering
	\includegraphics[width=\columnwidth]{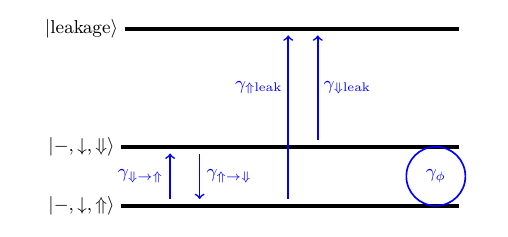}
	\caption{Effective three level decoherence model involving the two nuclear spin states, $\ket{-,\downarrow,\Uparrow}$ and $\ket{-,\downarrow,\Downarrow}$, and a leakage state $\ket{\mathrm{leakage}}$. The model allows for transitions between the two nuclear spin states with corresponding rates $\gamma_{\Downarrow\rightarrow\Uparrow}$ and $\gamma_{\Uparrow\rightarrow\Downarrow}$. Moreover, there is leakage of population from the nuclear spin subspace to the leakage state with nuclear spin dependent rates $\gamma_{\Uparrow\mathrm{leak}}$ and $\gamma_{\Downarrow\mathrm{leak}}$. Furthermore the model includes the dephasing between the nuclear spin states determined by the dephasing rate $\gamma_{\phi}$. All the rates are extracted from the decoherent time evolution determined by the master equation \eqref{eq:rhodotR} for initial density matrices restricted to the nuclear spin subspace.}
	\label{fig:effectivedecoherencemodel_single_system}
\end{figure}
 

We note that switching the drive on $(\epsilon_d\neq0)$ and off $(\epsilon_d=0)$ corresponds to switching the nuclear spin-photon coupling on and off. Therefore, implementing the $\sqrt{i\mathrm{SWAP}}$  or the $i\mathrm{SWAP}$ gate corresponds to driving both QDD systems for the associated gate time. That, in turn, implies that the qubit states are given by the two lowest energy eigenstates of $H_{\mathrm{QDD}}$~\eqref{eq:HQDD}. It is straightforward to verify that the qubit states are indeed the nuclear spin states $\ket{-,\downarrow,\Uparrow}$ and $\ket{-,\downarrow,\Uparrow}$. A transformation of the qubit states to the rotating reference frame yields 
\begin{align}
U_{\mathrm{ R,sys}}(t)\ket{-,\downarrow,\Uparrow(\Downarrow)}=e^{-i\omega_d t}\ket{-,\downarrow,\Uparrow(\Downarrow)}.
\end{align}
Since both states acquire the same phase independent of the nuclear spin orientation, all results obtained in the rotating reference frame for the nuclear spin subspace are also valid in the lab frame.

\begin{figure*}
     \centering
     \includegraphics[width=\textwidth]{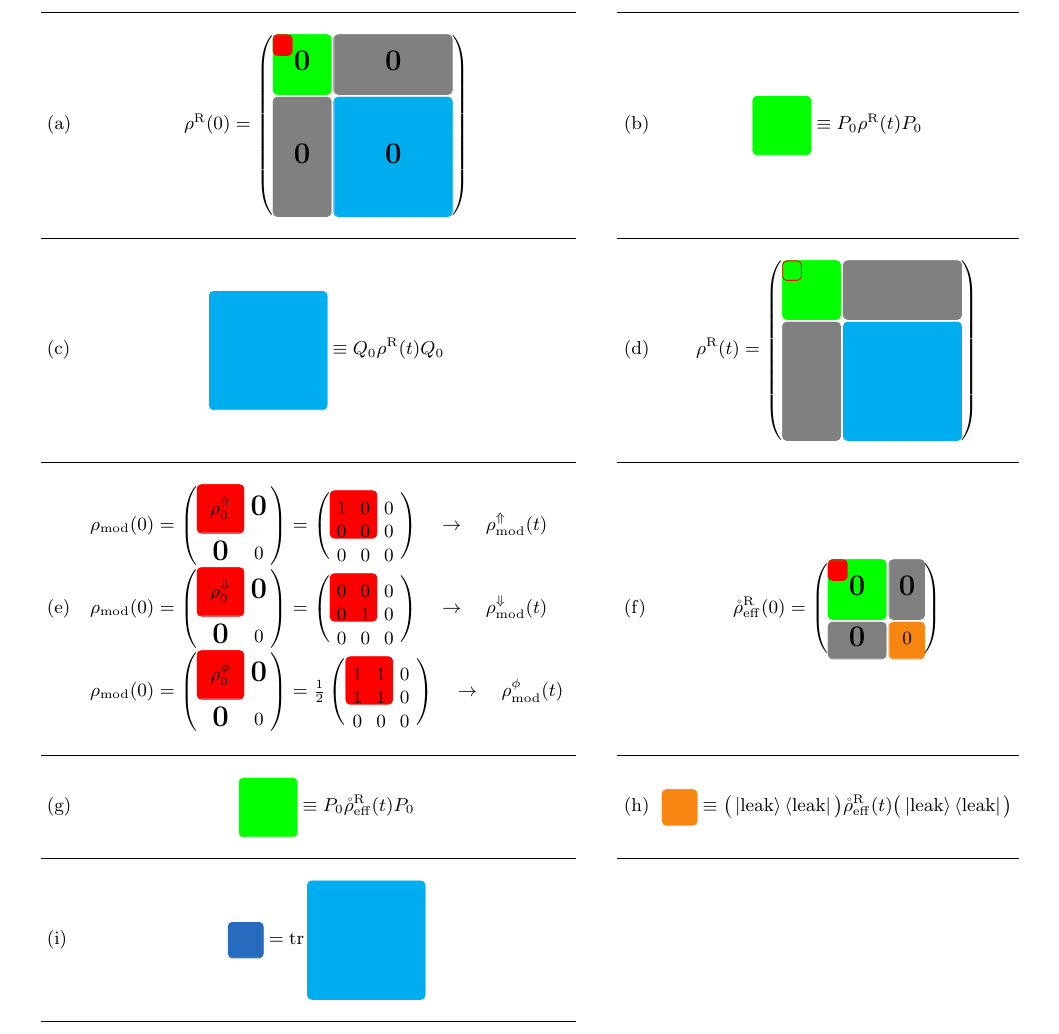}
     \caption{Schematic illustration and definition of the various density matrices and their individual blocks utilized in the effective decoherence model outlined in this appendix (App.~\ref{app:effective decoherence dynamics nuclear spin gate }).}
    \label{fig:coloredequations}
 \end{figure*}

Following the line of argument outlined in Appendix~\ref{app:nuclear_spin_decoherence_rate}, we find that the decoherence dynamics of the QDD system interacting with the resonator mode in the rotating reference frame are described by the Lindblad master equation 
\begin{align}
	\dot{\rho}^{\mathrm{ R}}(t)=&-i\left[H^{\mathrm{R}},\rho^{\mathrm{ R}}(t)\right]\nonumber\\
	&+\gamma^{\tau}\mathcal{D}\left[\tau_{-} \right]\rho^{\mathrm{ R}}(t)+\frac{\gamma_\phi^{\tau}}{2}\mathcal{D}\left[\tau_{z}\right]\rho^{\mathrm{ R}}(t)\nonumber\\
	&+\gamma^{\sigma}\mathcal{D}\left[\sigma_{-} \right]\rho^{\mathrm{ R}}(t)+\frac{\gamma_\phi^{\sigma}}{2}\mathcal{D}\left[	\sigma_{z}\right]\rho^{\mathrm{ R}}(t)\nonumber\\
	&+\kappa\mathcal{D}\left[a\right]\rho^{\mathrm{ R}}(t),
	\label{eq:rhodotR}
\end{align}
that, in addition to \eqref{eq:rhodotRsys} also accounts for the decay of resonator photons with rate $\kappa$. Within the scope of this work we are interested in the decoherence dynamics of initial states prepared in the qubit subspace once the drive and, therefore, the dispersive nuclear spin-photon coupling is switched on. We label the initial density matrix restricted to the qubit subspace and zero resonator photons, i.e. the space spanned by the set $\left\{\ket{-,\downarrow,\Uparrow,0}\ket{-,\downarrow,\Downarrow,0}\right\}$, by $\rho_{0}$ and depict it with a red square. For the following discussion the three specific choices for $\rho_{0}$ listed hereafter are of particular relevance:
\begin{align}
	\rho_{0}^{\Uparrow}&=\ket{-,\downarrow,\Uparrow}\bra{-,\downarrow,\Uparrow}
	=\begin{pmatrix}
		1 & 0 \\
		0 & 0
		\end{pmatrix},
	\label{eq:rho0up}\\
	\rho_{0}^{\Downarrow}&=\ket{-,\downarrow,\Downarrow}\bra{-,\downarrow,\Downarrow}
	=\begin{pmatrix}
		0 & 0 \\
		0 & 1
	\end{pmatrix},\\
\rho_{0}^{\phi}&=\left(\frac{\ket{-,\downarrow,\Uparrow}+\textcolor{Black}{\ket{-,\downarrow,\Downarrow}}}{\sqrt{2}}\right)\left(\frac{\bra{-,\downarrow,\Uparrow}+\textcolor{Black}{\bra{-,\downarrow,\Downarrow}}}{\sqrt{2}}\right)\nonumber\\
&=\frac{1}{2}\begin{pmatrix}
	1 & 1 \\
	1 & 1
\end{pmatrix}.
\label{eq:rho0phi}
\end{align}
Given $\rho_{0}$, the initial density matrix taking into account the entire space $\rho^{\mathrm{R}}(t=0)$ can be schematically represented in block form \textcolor{Black}{as shown in Fig.~\ref{fig:coloredequations}a,}
where all entries except the ones in the red block ($\equiv\rho_{0}$), that is part of the bigger green block, are vanishing. The green (blue) block 
\textcolor{Black}{ are defined in Fig.~\ref{fig:coloredequations}b(c).}
\textcolor{Black}{According to the defintions the green and blue block correspond to} the subspace spanned by the set $\{\ket{-,\downarrow,\Uparrow,n},\ket{-,\downarrow,\Downarrow,n}|n=0,1,2,...\}$,
and 
the subspace spanned by the set
$\{\ket{-,\uparrow,i,n},\ket{+,s,i,n}|n=0,1,2,...;\, s=\uparrow,\downarrow;\,i=\Uparrow,\Downarrow\}$
, respectively, while the gray blocks give the coherences between the green and the blue block. The time evolution of the initial density matrix $\rho^{\mathrm{ R}}(t)$ can be determined by solving the master equation \eqref{eq:rhodotR} numerically. Again, we represent the corresponding density matrix in block form \textcolor{Black}{as illustrated in Fig.~\ref{fig:coloredequations}d,}
where, in general, all the blocks have non vanishing entries. Within the scope of the dispersive two-qubit nuclear spin gate, the nuclear spin-photon coupling is tuned to the dispersive regime. Furthermore also the applied drive is off-resonant with the charge transition $(\ket{-}\leftrightarrow \ket{+})$. Thus, we expect that there is only a small population transfer from the subspace framed in red to the remainder of the green block and to the blue block.

Therefore, we expect that the dynamics of the nuclear spin qubit subspace can be effectively modelled by a three level model involving the two qubit states, $\ket{\Uparrow}$ and $\ket{\Downarrow}$, as well as a leakage state $\ket{\mathrm{leak}}$. Within this model we allow for population transfer between the nuclear spin qubit states with rates $\gamma_{\Downarrow \rightarrow \Uparrow}$ and  $\gamma_{\Uparrow \rightarrow \Downarrow}$, dephasing between the qubit states with corresponding rate $\gamma_{\phi}$, as well as population transfer from both qubit states to the leakage state characterized by the rates $\gamma_{\Downarrow \rightarrow \mathrm{leak}}$ and $\gamma_{\Uparrow \rightarrow \mathrm{leak}}$, as illustrated in Fig.~\ref{fig:effectivedecoherencemodel_single_system}. Assuming a diagonal Hamiltonian defined by the energies $E_{\Uparrow}$, $E_{\Downarrow}$, $E_{\mathrm{leak}}$, the decoherence dynamics of the three level model with respect to the basis $\{\ket{\Uparrow},\ket{\Downarrow},\ket{\mathrm{leak}}\}$ are described by the master equation 
\begin{align}
	\dot{\rho}_{\mathrm{mod}}(t)=&-i\left[\begin{pmatrix}
		E_{\Uparrow} & 0&0 \\
		0 & E_{\Downarrow} & 0 \\
		0 & 0 & E_{\mathrm{leak}}
		\end{pmatrix},\rho_{\mathrm{mod}}(t)\right ]\nonumber \\
	&+\gamma_{\Downarrow \rightarrow \Uparrow}\mathcal{D}\left[\begin{pmatrix}
		0 & 1 & 0 \\
			0 & 0 & 0 \\
				0 & 0 & 0 \\
		\end{pmatrix}\right]\rho_{\mathrm{mod}}(t)\nonumber\\
	&+\gamma_{\Uparrow \rightarrow \Downarrow}\mathcal{D}\left[\begin{pmatrix}
		0 & 0 & 0 \\
		1& 0 & 0 \\
		0 & 0 & 0 \\
	\end{pmatrix}\right]\rho_{\mathrm{mod}}(t)\nonumber\\
&+\frac{\gamma_{\phi}}{2}\mathcal{D}\left[\begin{pmatrix}
	-1& 0 & 0 \\
	0& 1 & 0 \\
	0 & 0 & 0 \\
\end{pmatrix}\right]\rho_{\mathrm{mod}}(t)\nonumber\\
&+\gamma_{\Uparrow \rightarrow \mathrm{leak}}\mathcal{D}\left[\begin{pmatrix}
0 & 0 & 0 \\
0 & 0 & 0 \\
1& 0 & 0 \\
\end{pmatrix}\right]\rho_{\mathrm{mod}}(t)\nonumber\\
&+\gamma_{\Downarrow \rightarrow \mathrm{leak}}\mathcal{D}\left[\begin{pmatrix}
	0 & 0 & 0 \\
	0 & 0 & 0 \\
	0& 1 & 0 \\
\end{pmatrix}\right]\rho_{\mathrm{mod}}(t).
\label{eq:rhomoddot}
\end{align}
The above master equation can be solved analytically and, thus, allows to find analytical expressions for the time evolution of the system initially prepared in each of the qubit states and also for the equal superposition of the two qubit states. The resulting density matrices after the decoherent time evolution are labeled in the 
way \textcolor{Black}{defined in Fig.~\ref{fig:coloredequations}e.}

Aiming at finding the effective decoherence rates that allow to model the dynamics of the full system with the three level model with high accuracy, we first describe how the different rates can be obtained from the solutions \textcolor{Black}{$\rho^{\Downarrow}_{\mathrm{mod}}(t)$, $\rho^{\Uparrow}_{\mathrm{mod}}(t)$, and $\quad \rho^{\phi}_{\mathrm{mod}}(t)$}
\textcolor{Black}{(see Fig.~\ref{fig:coloredequations}e)} of the master equation \eqref{eq:rhomoddot} and later apply these findings to obtain the effective rates.

First, we find that the ratio $c_{\gamma}=\gamma_{\Uparrow \rightarrow \Downarrow}/\gamma_{\Downarrow \rightarrow \Uparrow}$ can be extracted from the time evolved density matrices for different initial states
\begin{align}
	\frac{ \rho^{\Uparrow}_{\mathrm{mod};\Downarrow,\Downarrow}(t)}{\rho^{\Downarrow}_{\mathrm{mod};\Uparrow,\Uparrow}(t) }
=\frac{\gamma_{\Uparrow \rightarrow \Downarrow}}{\gamma_{\Downarrow \rightarrow \Uparrow}}=c_{\gamma} \text{ for }t\neq 0,
\label{eq:cgamma}
\end{align}
where  $\rho^{\beta}_{\mathrm{mod};i,j}(t)$ with $\beta\in\{\Uparrow,\Downarrow,\phi\}$ and $i,j\in\{\Uparrow,\Downarrow,\mathrm{leak}\}$  labels the matrix element of the respective density matrix:
\begin{align}
	\rho^{\beta}_{\mathrm{mod}}(t) =	 \sum_{i,j}\rho^{\beta}_{\mathrm{mod};i,j}(t)\ket{i}\bra{j}.
\end{align}
In addition, the density matrices also provide information about the ratio between the difference of the leakage rates and $\gamma_{\Uparrow \rightarrow \Downarrow}$
\begin{align}
	&\frac{\rho^{\Uparrow}_{\mathrm{mod};\mathrm{leak, leak}}(t)-\rho^{\Downarrow}_{\mathrm{mod};\mathrm{leak,leak}}(t)}{\rho^{\Uparrow}_{\mathrm{mod};\Downarrow,\Downarrow}(t)}=\frac{\gamma_{\Uparrow\rightarrow\mathrm{leak}}-\gamma_{\Downarrow\rightarrow\mathrm{leak}}}{\gamma_{\Uparrow \rightarrow \Downarrow}}\nonumber\\
	&\quad\quad=\frac{\gamma_{\mathrm{leak-diff}}}{\gamma_{\Uparrow \rightarrow \Downarrow}}=c_{\gamma\mathrm{leak}} \text{ for }t\neq 0.
	\label{eq:cgammaleak}
\end{align}
Furthermore, we find that 
\begin{widetext}
\begin{align}
	\frac{ \rho^{\Uparrow}_{\mathrm{mod};\Uparrow,\Uparrow}(t)}{\rho^{\Downarrow}_{\mathrm{mod};\Downarrow,\Downarrow}(t) }=\frac{c_{\gamma}\left(1+e^{tb}\right)b-\left(-1+c_{\gamma}+c_{\gamma}c_{\gamma\mathrm{leak}}\right)\left(-1+e^{tb}\right)\gamma_{\Uparrow\rightarrow\Downarrow}}
	{c_{\gamma}\left(1+e^{tb}\right)b+\left(-1+c_{\gamma}+c_{\gamma}c_{\gamma\mathrm{leak}}\right)\left(-1+e^{tb}\right)\gamma_{\Uparrow\rightarrow\Downarrow}},
	\label{eq:gamnmaupdowndetermine}
\end{align}
with
\begin{align}
	b=\sqrt{\frac{\gamma_{\Uparrow\rightarrow\Downarrow}^2 \left[c_{\gamma}\left(c_{\gamma}[c_{\gamma\mathrm{leak}}+1]^2-2 c_{\gamma\mathrm{leak}}+2\right)+1\right]}{c_{\gamma}^2}} ,
\end{align}
as a function of time is fully determined by the parameters $c_{\gamma}$, $c_{\gamma\mathrm{leak}}$ and $\gamma_{\Uparrow\rightarrow\Downarrow}$ of which the first two are already determined by  \eqref{eq:cgamma} and \eqref{eq:cgammaleak}, such that using the right hand side of \eqref{eq:gamnmaupdowndetermine} as a fit function for the time evolution of the ratio $ \rho^{\Uparrow}_{\mathrm{mod};\Uparrow,\Uparrow}(t)/\rho^{\Downarrow}_{\mathrm{mod};\Downarrow,\Downarrow}(t)$ with fitting parameter $\gamma_{\Uparrow\rightarrow\Downarrow}$ yields the value of $\gamma_{\Uparrow\rightarrow\Downarrow}$. Thereby, also $\gamma_{\Downarrow \rightarrow \Uparrow}$ is determined as $\gamma_{\Downarrow \rightarrow \Uparrow}=\gamma_{\Uparrow\rightarrow\Downarrow}/c_{\gamma}$ according to \eqref{eq:cgamma}. 
As a next step we note that the analytical form  of the matrix element $\rho^{\Uparrow}_{\mathrm{mod};\Downarrow,\Downarrow}(t)$ reads
\begin{align}
\rho^{\Uparrow}_{\mathrm{mod};\Downarrow,\Downarrow}(t)=\frac{e^{-t\frac{\gamma_{\Uparrow\rightarrow\Downarrow}+c_{\gamma}\left(d+2\gamma_{\Downarrow\rightarrow\mathrm{leak}}+\gamma_{\Uparrow\rightarrow\Downarrow}+c_{\gamma\mathrm{leak}}\gamma_{\Uparrow\rightarrow\Downarrow}\right)}{2 c_{\gamma}}}\left(-1+e^{td}\right)\gamma_{\Uparrow\rightarrow\Downarrow}}{d},
\label{eq:rhoupfit}
\end{align}
with
\begin{align}
	d=\sqrt{\gamma_{\Uparrow\rightarrow\Downarrow}^2\frac{1-2c_{\gamma}\left(-1+c_{\gamma\mathrm{leak}}\right)+c_{\gamma}^2\left(1+c_{\gamma\mathrm{leak}}\right)^2}{c_{\gamma}^2}}.
\end{align}
Obviously, following the previous discussion, $\gamma_{\Downarrow\rightarrow \mathrm{leak}}$ is the only undetermined parameter defining the time evolution of the matrix element. Thus, using the right hand side of \eqref{eq:rhoupfit} as a fit function for $\rho^{\Uparrow}_{\mathrm{mod};\Downarrow,\Downarrow}(t)$ allows to determine $\gamma_{\Downarrow\rightarrow \mathrm{leak}}$. At the same time also $\gamma_{\Uparrow\rightarrow \mathrm{leak}}=c_{\gamma\mathrm{leak}}\gamma_{\Uparrow\rightarrow\Downarrow}+\gamma_{\Downarrow\rightarrow \mathrm{leak}}$ is specified because $c_{\gamma\mathrm{leak}}$ and $\gamma_{\Uparrow\rightarrow\Downarrow}$ have been obtained before. In a similar fashion $\gamma_{\phi}$ can be obtained from a fit of	$\rho^{\phi}_{\mathrm{mod};\Uparrow,\Downarrow}(t)$ because 
 \begin{align}
	\rho^{\phi}_{\mathrm{mod};\Uparrow,\Downarrow}(t)= \frac{1}{2}e^{it\left(E_{\Downarrow}-E_{\Uparrow}\right)}e^{-t\frac{\gamma_{\Uparrow\rightarrow\Downarrow}+\gamma_{\Downarrow}\rightarrow\Uparrow+\gamma_{\Uparrow \rightarrow \mathrm{leak}}+\gamma_{\Downarrow \rightarrow \mathrm{leak}}+2\gamma_{\phi}}{2}} 
\end{align}
and, in particular, 
 \begin{align}
	\left|\rho^{\phi}_{\mathrm{mod};\Uparrow,\Downarrow}(t)\right|= \frac{1}{2}e^{-t\frac{\gamma_{\Uparrow\rightarrow\Downarrow}+\gamma_{\Downarrow}\rightarrow\Uparrow+\gamma_{\Uparrow \rightarrow \mathrm{leak}}+\gamma_{\Downarrow \rightarrow \mathrm{leak}}+2\gamma_{\phi}}{2}},
\end{align}
holds. 

As the next step we numerically solve the master equation \eqref{eq:rhodotR} for initial density matrices of the form 
\textcolor{Black}{given in Fig.}~\ref{fig:coloredequations}a with $\rho_{0}=\rho_{0}^{\Uparrow},\rho_{0}^{\Downarrow},\rho_{0}^{\phi}$ and denote the resulting density matrices by $\rho^{\mathrm{ R},\Uparrow}(t),\rho^{\mathrm{ R},\Downarrow}(t)$ and $\rho^{\mathrm{ R},\phi}(t)$, respectively. Any density matrix $\rho^{R}(t)$, and, in particular, the three just mentioned ones, can be written as 
\begin{align}
	\rho^{R}(t)= \sum_{\substack{k,k',l,l'\\m,m',n,n'}}\rho^{\mathrm{ R}}_{(k,l,m,n),(k',l',m',n')}(t) \ket{k,l,m,n}\bra{k',l',m',n'}\\\
	\text{    with    }\,
	k,k'\in\{-,+\};\,l,l'\in\{\downarrow,\uparrow\};\, m,m'\in\{\Uparrow,\Downarrow\};\, n,n'=0,1,2,...\,,\nonumber
\end{align}
and the matrix elements $\rho^{\mathrm{ R}}_{(k,l,m,n),(k',l',m',n')}(t)$ with respect to the chosen basis. 

In order to extract the effective decoherence rates defining the three level model from the numerically obtained solutions for $\rho^{\mathrm{R}}(t)$ we identify specific matrix elements of the three level model with matrix elements of the full system in the following way:
\begin{align}
	\begin{array}{lcl}
		 \rho^{\Uparrow(\Downarrow)}_{\mathrm{mod};\Downarrow,\Downarrow}(t)&\rightarrow &\rho^{\mathrm{ R},\Uparrow(\Downarrow)}_{(-,\downarrow,\Downarrow,0),(-,\downarrow,\Downarrow,0)}(t),\\
		 &&\\
	\rho^{\Uparrow(\Downarrow)}_{\mathrm{mod};\Uparrow,\Uparrow}(t) &\rightarrow & \rho^{\mathrm{ R},\Uparrow(\Downarrow)}_{(-,\downarrow,\Uparrow,0),(-,\downarrow,\Uparrow,0)}(t),\\
	&&\\
      \rho^{\Uparrow(\Downarrow)}_{\mathrm{mod};\mathrm{leak, leak}}(t)&\rightarrow &\sum_{\substack{
				k,l,m,n 	\\
				(k,l,n)\neq(-,\downarrow,0)
		}}\rho^{\mathrm{ R},\Uparrow(\Downarrow)}_{(k,l,m,n),(k,l,m,n)}(t),\\
	&&\\
		\rho^{\phi}_{\mathrm{mod};\Uparrow,\Downarrow}(t)&\rightarrow &\rho^{\mathrm{ R},\phi}_{(-,\downarrow,\Uparrow,0),(-,\downarrow,\Downarrow,0)}(t).
		\end{array}
\end{align}
\end{widetext}
Using this identification, we follow the detailed discussion outlined earlier in this appendix and determine the effective decoherence rates.

\begin{figure*}[!ht]
 \centering
\includegraphics[width=\textwidth]{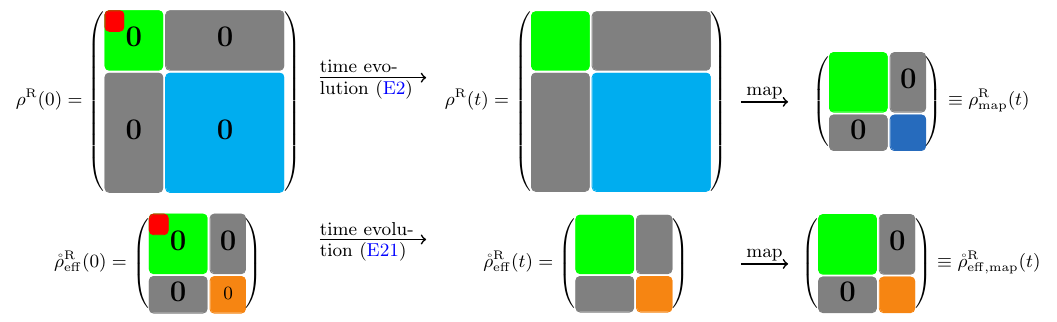}
\caption{Schematic representation of the mapping that allows to compare the dynamics of the full system with those of the effective model. First, the initial density matrices $\rho^{\mathrm{ R}}(0)$ and $\mathring{\rho}^{\mathrm{R}}_{\mathrm{eff}}(0)$ are both determined by a given $\rho_0$ (red square). Then, the initial states evolve according to the Lindblad master equations \eqref{eq:rhodotR}  and \eqref{eq:rhoringeffdot}, respectively. As a next step, the light blue block is mapped to the one-dimensional dark blue block, whereby the entry of the latter is given by the trace of the light blue block (see \textcolor{Black}{Fig.~\ref{fig:coloredequations}i}).
In addition, the coherences (gray blocks) between the green and the dark blue as well as the \textcolor{Black}{coherences} between the green and the orange block are set to zero. The exact definitions of the various blocks are given in Appendix~\ref{app:eff_decoherence_single_QDD}.}
	\label{fig:mappingdensitymatrices}
  \end{figure*}

We can now compare the decoherence dynamics generated by the effective low dimensional system with those of the full system for initial density matrices restricted to the nuclear spin subspace and an empty resonator in order to check the accuracy of the effective model. In Sec.~\ref{sec:nuclear spin-photon coupling} we have derived the effective nuclear spin-photon Hamiltonian $H_{\mathrm{eff}}^{\mathrm{ R}}$~\eqref{eq:Heffsingle}. However, for the numerical analysis the effective Hamiltonian $\mathring{H}_{\mathrm{eff}}^{\mathrm{ R}}$ is used to increase the accuracy. The derivation of this effective Hamiltonian involves an additional step compared to the derivation of  $H_{\mathrm{eff}}$.  As the additional step we find the unitary $U_{\mathrm{QDD}}$ that diagonalizes $H_{\mathrm{QDD}}$ transformed to the rotating reference frame, i.e. 
\begin{align*}
U_{\mathrm{QDD}}(t)&\left( U_{\mathrm{ R},sys}(t)H_{\mathrm{QDD}}U^{\dagger}_{\mathrm{ R},sys}(t)\right.\\
&\left.+i\dot{U}_{\mathrm{ R},sys}(t)U^{\dagger}_{\mathrm{ R},sys}(t)   \right)U^{\dagger}_{\mathrm{QDD}}(t)
\end{align*}
is a diagonal operator. Here, it is important to point out that the nuclear spin qubit states $\ket{-,\downarrow,\Uparrow}$ and $\ket{-,\downarrow,\Downarrow}$ are eigenstates of $H_{\mathrm{QDD}}$ transformed to the rotating reference frame, and, therefore, constitute two of the basis states of the basis set by $U_{\mathrm{QDD}}$. Thus, the projection operator $P_{0}$~\eqref{eq:P0} is still identifying the subspace whose dynamics we want to describe with an effective Hamiltonian. 
The advantage of the transformation to the basis set by $U_{\mathrm{QDD}}$ lies in the fact that the magnetic field gradient (first term in $V^{\mathrm{ R}}$~\eqref{eq:VR}) and the hyperfine interaction (third term in $V^{\mathrm{ R}}$~\eqref{eq:VR}) are no longer treated as a perturbation when applying a Schrieffer-Wolff transformation in the following way: First, $H^{\mathrm{ R}}$ is transformed to the basis defined by $U_{\mathrm{QDD}}$,
\begin{align}
	\mathring{H}^{\mathrm{ R}}=U_{\mathrm{QDD}} H^{\mathrm{R}} U_{\mathrm{QDD}}^{\dagger}, 
\end{align}
then $\mathring{H}^{\mathrm{ R}}$ is split in a diagonal contribution $\mathring{H}^{\mathrm{ R}}_0$ and an off-diagonal part $\mathring{V}^{\mathrm{ R}}$. The two parts are defined as 
\begin{align}
\mathring{H}^{\mathrm{ R}}_0=	\sum_{s=1}^{8}\sum_{n=0}^{\infty}\bra{s,n}\mathring{H}^{\mathrm{ R}}\ket{s,n}\ket{s,n}\bra{s,n},
\end{align}
where the first entry of $\ket{s,n}$ labels the eight states of the basis defined by $U_{\mathrm{QDD}}$ while $n$ gives the resonator photon number, and
\begin{align}
	\mathring{V}^{\mathrm{ R}}=	\mathring{H}^{\mathrm{ R}}-\mathring{H}^{\mathrm{ R}}_0.
\end{align}
Then, we follow the discussion presented in Appendix~\ref{app:eff_nuclear_spin_photon_coupling} by using the replacements $H_0^{\mathrm{ R}}\rightarrow \mathring{H}_0^{\mathrm{ R}}$  and  $V^{\mathrm{ R}}\rightarrow \mathring{V}^{\mathrm{ R}}$ to numerically determine the effective Hamiltonian $\mathring{H}^{\mathrm{ R}}_{\mathrm{eff}}$ up to sixth order in the perturbation $\mathring{V}^{\mathrm{ R}}$. Thereby the photon number space is restricted to 0 to 6 photons to allow the numerical treatment. 

At this point, an effective master equation that is expected to reproduce the decoherence dynamics of the nuclear spin subspace interacting with the cavity mode with high accuracy if the system is initialized in the nuclear spin subspace and an empty resonator can be set up,
\begin{align}
	\dot{\mathring{\rho}}^{\mathrm{R}}_{\mathrm{eff}}(t)=&-i\left[\mathring{H}_{\mathrm{eff}}\oplus E_{\mathrm{leak}}\ket{\mathrm{leak}}\bra{\mathrm{leak}}
,	\mathring{\rho}^{\mathrm{R}}_{\mathrm{eff}}(t)\right ]\nonumber \\
	&+\gamma_{\Downarrow \rightarrow \Uparrow}\mathcal{D}\left[\nu_+\right]	\mathring{\rho}^{\mathrm{R}}_{\mathrm{eff}}(t)\nonumber\\
	&+\gamma_{\Uparrow \rightarrow \Downarrow}\mathcal{D}\left[\nu_-\right]	\mathring{\rho}^{\mathrm{R}}_{\mathrm{eff}}(t)\nonumber\\
	&+\frac{\gamma_{\phi}}{2}\mathcal{D}\left[\nu_z\right]	\mathring{\rho}^{\mathrm{R}}_{\mathrm{eff}}(t)\nonumber\\
	&+\gamma_{\Uparrow \rightarrow \mathrm{leak}}\mathcal{D}\left[\sum_{n=0}^{\infty}\ket{\mathrm{leak}}\bra{\Uparrow,n}\right]	\mathring{\rho}^{\mathrm{R}}_{\mathrm{eff}}(t)\nonumber\\
	&+\gamma_{\Downarrow \rightarrow \mathrm{leak}}\mathcal{D}\left[\sum_{n=0}^{\infty}\ket{\mathrm{leak}}\bra{\Downarrow,n}\right]	\mathring{\rho}^{\mathrm{R}}_{\mathrm{eff}}(t),
	\label{eq:rhoringeffdot}
\end{align}
where we have introduced the energy of the leakage state $E_{\mathrm{leak}}$. However, since the following analysis does not involve the coherences between the leakage state and the rest of the Hilbert space, the value of $E_{\mathrm{leak}}$ can be chosen arbitrarily and we choose to set it to 0 in our simulations. 

In analogy to 
\textcolor{Black}{Fig.~\ref{fig:coloredequations}a} the relevant initial density matrices within the scope of this work can schematically be written as \textcolor{Black}{illustrated in Fig.~\ref{fig:coloredequations}f,}
where the green block \textcolor{Black}{defined in Fig.~\ref{fig:coloredequations}g} represents the same subspace as its counterpart in 
\textcolor{Black}{Fig.~\ref{fig:coloredequations}a}, 
and the one-dimensional leakage subspace is illustrated by the orange block \textcolor{Black}{specified in Fig.~\ref{fig:coloredequations}h.} 

It remains to check whether \eqref{eq:rhoringeffdot} indeed reproduces the decoherence dynamics generated by \eqref{eq:rhodotR} reliably. In order to do that the density matrices $\rho^{\mathrm{R}}(t)$ and $\mathring{\rho}^{\mathrm{R}}_{\mathrm{eff}}(t)$ obtained for $\rho^{\mathrm{R}}(0)$ and $\mathring{\rho}^{\mathrm{R}}_{\mathrm{eff}}(0)$  both determined by the same $\rho_{0}$ have to be compared in some way. Since $\rho^{\mathrm{R}}(t)$ and $\mathring{\rho}^{\mathrm{R}}_{\mathrm{eff}}(t) $ have different dimension there is no straightforward way to calculate the trace distance or the fidelity between the two states. This work focuses on the interaction dynamics between the nuclear spin qubit states and the microwave resonator, i.e. the green block in the schematic representation of the density matrices, while only the total population but not the exact form of the blue block in 
\textcolor{Black}{Fig.~\ref{fig:coloredequations}d} matters. Besides that also the coherences (gray blocks) between the green and the blue block of $\rho^{\mathrm{R}}(t)$ as well as the green and the orange block of  $\mathring{\rho}^{\mathrm{R}}_{\mathrm{eff}}(t) $ are not relevant here. Following this line of argument we introduce the mapping presented in Fig.~\ref{fig:mappingdensitymatrices} that maps $\rho^{\mathrm{R}}(t)$ and $\mathring{\rho}^{\mathrm{R}}_{\mathrm{eff}}(t) $ to the matrices  $\rho^{\mathrm{ R}}_{\mathrm{map}}(t)$ and $\mathring{\rho}^{\mathrm{ R}}_{\mathrm{eff,map}}(t)$ of the same dimension with trace 1. The latter property is ensured by introducing the dark blue block that corresponds to the sum of all populations of the blue block \textcolor{Black}{as defined in Fig.~\ref{fig:coloredequations}i.}  

Before finally comparing   $\rho^{\mathrm{ R}}_{\mathrm{map}}(t)$ and $\mathring{\rho}^{\mathrm{ R}}_{\mathrm{eff,map}}(t)$, we note that $\rho_{0}$ is the density matrix of a nuclear spin qubit and, therefore, any possible  $\rho_0$ can be written as
\begin{align}
	\rho_0=\frac{1+\vec{a}\cdot \vec{\nu}}{2} \text{ with }|\vec{a}|\leq1,
\end{align}
characterized by the Bloch vector $\vec{a}$. Then the fidelity  between the quantum states $\rho^{\mathrm{ R}}_{\mathrm{map}}(t)$ and $\mathring{\rho}^{\mathrm{ R}}_{\mathrm{eff,map}}(t)$ for a given initial state characterized by the Bloch vector $\vec{a}$ and after decoherent time evolution from 0 to $t$ reads \cite{nielsen2010}
\begin{align}
	F\left(\vec{a},t\right)=\tr\sqrt{\left[\rho^{\mathrm{ R}}_{\mathrm{map}}(\vec{a},t)\right]^{\tfrac{1}{2}}\mathring{\rho}^{\mathrm{ R}}_{\mathrm{eff,map}}(\vec{a},t)\left[\rho^{\mathrm{ R}}_{\mathrm{map}}(\vec{a},t)\right]^{\tfrac{1}{2}}}.
\end{align}
 However, we want the effective model to model the decoherent time evolution of any initial state characterized by $\rho_0$ accurately, wherefore we define the average fidelity as 
 \begin{align}
	\bar{F}\left(t\right)=\frac{1}{\tfrac{4}{3}\pi}\int_{0}^{1}\int_{0}^{2\pi} \int_{0}^{\pi}F\left(\vec{a},t\right)a^2\sin(\theta)\,da\,d\phi\, d\theta,
	\label{eq:Favg}
\end{align}
with $\vec{a}$ expressed in spherical coordinates, $\vec{a}=|\vec{a}|\left(\cos(\phi)\sin(\theta),\sin(\phi)\sin(\theta),\cos(\theta)\right)$, and use this quantity as a measure for the accuracy of the introduced effective model.

\begin{figure}
	\centering
	\includegraphics[width=\columnwidth]{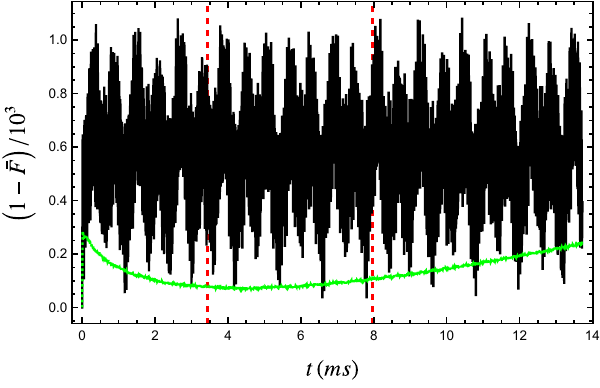}
	\caption{Average infidelity \eqref{eq:Favg} of the states $\rho^{\mathrm{ R}}_{\mathrm{map}}(t)$ and $\mathring{\rho}^{\mathrm{ R}}_{\mathrm{eff,map}}(t)$ as a function of time for coherent evolution (black) and evolution taking into account decoherence effects (green). The infidelity is far below 1\% on the considered timescale that exceeds both the gate time for the $\sqrt{i\mathrm{SWAP}}$ gate (left vertical red dashed line) and the gate time for the $i\mathrm{SWAP}$ gate (right vertical red dashed line). These gate times are obtained from numerical simulations as described in Appendix~\ref{app:effective decoherence dynamics nuclear spin gate }. \textcolor{Black}{For the decoherence model used for this plot we assume a cavity quality factor of $Q=10^5$, $T_{1}^{\tau}=T_{2}^{\tau}/2$  and the decay and decoherence times $T_{2}^{\tau},T_{1}^{\sigma}$ and $T_{2}^{\sigma}$ listed in Appendix~\ref{app:nuclear_spin_decoherence_rate}.} The system parameters agree with those given in Appendix~\ref{app:effective decoherence dynamics nuclear spin gate }.}
	\label{fig:average_fidelity_effective_model}
\end{figure}

Figure~\ref{fig:average_fidelity_effective_model} shows the average infidelity, $1-\bar{F}$, for the coherent time evolution in black and the decoherent time evolution in green. In both cases the infidelity is far below $1\%$ for the considered timescale that exceeds the gate time of the $\sqrt{i\mathrm{SWAP}}$ and $i\mathrm{SWAP}$ gate. Therefore, we conclude that the effective decoherence model indeed reproduced the decoherence dynamics of the full system for initial states defined by $\rho_0$ on timescales relevant for this work reliably. 

\subsection{Effective decoherence dynamics nuclear spin gate \label{app:effective decoherence dynamics nuclear spin gate }}
As a next step the effective model for the decoherence dynamics of a single nuclear spin qubit realized in a QDD system interacting with a microwave resonator is employed to predict the decoherence dynamics of the effective interaction between two nuclear spin qubits mediated by the microwave resonator and, thus, ultimately allows to obtain estimates for the fidelity of $\sqrt{i\mathrm{SWAP}}$ and the $i\mathrm{SWAP}$ gate. 

In line with the discussion Appendix~\ref{app:eff_decoherence_single_QDD}, we transform $\hat{H}^{\mathrm{ R}}$ to the basis that diagonalizes $H_{\mathrm{QDD}}^{(1)}$ and $H_{\mathrm{QDD}}^{(2)}$ transformed to the rotating reference,
\begin{align}
	\breve{H}^{\mathrm{ R}}=U_{\mathrm{QDD}}^{(1)} U_{\mathrm{QDD}}^{(2)}\hat{H}^{\mathrm{ R}}\left(U_{\mathrm{QDD}}^{(1)}U_{\mathrm{QDD}}^{(2)}\right)^{\dagger},
\end{align} 
to increase the accuracy of the effective Hamiltonian obtained by applying a Schrieffer-Wolff transformation subsequently. Again, the subspace defined by $\hat{P}_0$ is not affected by the basis transformation. With respect to the basis set by $U_{\mathrm{QDD}}^{(1)} U_{\mathrm{QDD}}^{(2)}$, the diagonal part of $\breve{H}^{\mathrm{ R}}$ reads 
\begin{align}
	\breve{H}^{\mathrm{ R}}_0=&	\sum_{s^{(1)},s^{(2)}=1}^{8}\sum_{n=0}^{\infty}\Bigg[\bra{s^{(1)},s^{(2)},n}\breve{H}^{\mathrm{ R}}\ket{s^{(1)},s^{(2)},n}\nonumber\\
	&\quad\quad\quad\quad\quad\quad\quad\quad\ket{s^{(1)},s^{(2)},n}\bra{s^{(1)},s^{(2)},n}\Bigg],
\end{align}
with $s^{(1)}$ and $s^{(2)}$ labeling the eight eigenstates of $H_{QDD}^{(1)}$ and $H_{QDD}^{(2)}$ transformed to the rotating frame, respectively. Then, the off-diagonal perturbation is given by 
\begin{align}
	\breve{V}^{\mathrm{ R}}=	\breve{H}^{\mathrm{ R}}-\breve{H}^{\mathrm{ R}}_0.
\end{align}
Using these two definitions, we follow the reasoning in Appendix~\ref{app:coupled_nuclear_spins} to numerically obtain the effective Hamiltonian $	\breve{H}^{\mathrm{R}}_{\nu-\mathrm{ph}}$ capturing the interaction of the two nuclear spin qubits interacting with the microwave resonator up to sixth order in the perturbation $\breve{V}^{\mathrm{ R}}$. Finally, we can set up the effective master equation for the two nuclear spin qubits interacting with the microwave resonator:
\begin{align}
	\dot{\breve{\rho}}^{\mathrm{ R}}_{\mathrm{\nu-\mathrm{ph}}}(t)=&-i\left[	\breve{H}^{\mathrm{R}}_{\nu-\mathrm{ph}}\oplus E_{\mathrm{leak}}\ket{\mathrm{leak}}\bra{\mathrm{leak}}
	,\breve{\rho}^{\mathrm{ R}}_{\mathrm{\nu-\mathrm{ph}}}(t)\right ]\nonumber \\
	&+\sum_{i=1}^2\bigg[\gamma_{\Downarrow \rightarrow \Uparrow}\mathcal{D}\left[\nu^{(i)}_+\right]\breve{\rho}^{\mathrm{ R}}_{\mathrm{\nu-\mathrm{ph}}}(t)\nonumber\\
	&\quad\quad\quad+\gamma_{\Uparrow \rightarrow \Downarrow}\mathcal{D}\left[\nu^{(i)}_-\right]\breve{\rho}^{\mathrm{ R}}_{\mathrm{\nu-\mathrm{ph}}}(t)\nonumber\\
	&\quad\quad\quad+\frac{\gamma_{\phi}}{2}\mathcal{D}\left[\nu_z^{(i)}\right]\breve{\rho}^{\mathrm{ R}}_{\mathrm{\nu-\mathrm{ph}}}(t)\bigg]\nonumber\\
	&+\gamma_{\Uparrow \rightarrow \mathrm{leak}}\mathcal{D}\Bigg[\sum_{n=0}^{\infty}\bigg(\ket{\mathrm{leak}}\bra{\Uparrow,\Uparrow,n}\nonumber\\
	&\quad\quad\quad\quad\quad\quad\quad+\ket{\mathrm{leak}}\bra{\Uparrow,\Downarrow,n}\bigg)\Bigg]\breve{\rho}^{\mathrm{ R}}_{\mathrm{\nu-\mathrm{ph}}}(t)\nonumber\\
	&+\gamma_{\Uparrow \rightarrow \mathrm{leak}}\mathcal{D}\Bigg[\sum_{n=0}^{\infty}\bigg(\ket{\mathrm{leak}}\bra{\Uparrow,\Uparrow,n}\nonumber\\
	&\quad\quad\quad\quad\quad\quad\quad+\ket{\mathrm{leak}}\bra{\Downarrow,\Uparrow,n}\bigg)\Bigg]\breve{\rho}^{\mathrm{ R}}_{\mathrm{\nu-\mathrm{ph}}}(t)\nonumber\\
	&+\gamma_{\Downarrow \rightarrow \mathrm{leak}}\mathcal{D}\Bigg[\sum_{n=0}^{\infty}\bigg(\ket{\mathrm{leak}}\bra{\Downarrow,\Uparrow,n}\nonumber\\
	&\quad\quad\quad\quad\quad\quad\quad+\ket{\mathrm{leak}}\bra{\Downarrow,\Downarrow,n}\big)\Bigg]\breve{\rho}^{\mathrm{ R}}_{\mathrm{\nu-\mathrm{ph}}}(t)\nonumber\\
	&+\gamma_{\Downarrow \rightarrow \mathrm{leak}}\mathcal{D}\Bigg[\sum_{n=0}^{\infty}\bigg(\ket{\mathrm{leak}}\bra{\Uparrow,\Downarrow,n}\nonumber\\
	&\quad\quad\quad\quad\quad\quad\quad+\ket{\mathrm{leak}}\bra{\Downarrow,\Downarrow,n}\bigg)\Bigg]\breve{\rho}^{\mathrm{ R}}_{\mathrm{\nu-\mathrm{ph}}}(t)\nonumber\\
	\label{eq:rhobrevenu-phdot}
\end{align}
In the dispersive nuclear spin-photon coupling regime we expect an effective interaction between the nuclear spin qubits mediated by virtual resonator photons, as discussed in detail in Sec.~\ref{sec:dispersive_nuclear_spin_gate}. In particular, the coherent time evolution can realize a $\sqrt{i\mathrm{SWAP}}$ and an $i\mathrm{SWAP}$  quantum gate. In order to assess decoherence effects on the gate fidelity, a closer investigation of the zero photon subspace of 
	$\breve{\rho}^{\mathrm{ R}}_{\mathrm{\nu-\mathrm{ph}}}(t)$ is necessary. This corresponds to the projection 
	\begin{align}
		\breve{\rho}^{\mathrm{R,sub}}_{\mathrm{\nu-\mathrm{ph}}}(t)=	P_{\mathrm{sub}}\breve{\rho}^{\mathrm{ R}}_{\mathrm{\nu-\mathrm{ph}}}(t)P_{\mathrm{sub}},
		\label{eq:rhonuphsub}
	\end{align}
with 
\begin{align}
	P_{\mathrm{sub}}=\sum_{\nu^{(1)},\nu^{(2)}\in\{\Uparrow,\Downarrow\}}\ket{\nu^{(1)},\nu^{(2)},0}\bra{\nu^{(1)},\nu^{(2)},0}.
\end{align}
Equation \eqref{eq:rhonuphsub} represents a trace-nonpreserving quantum map. This property has to be taken into account when calculating the average gate fidelity $\bar{F}_{\mathrm{gate}}$ for the system initialized in the subspace defined by $P_{\mathrm{sub}}$ because the usual approach to calculate the average fidelity \cite{horodecki1999,nielsen2002} requires a trace preserving quantum map.  However, this approach has recently been modified to allow for trace-nonpreserving quantum maps \cite{shkolnikov2020}, and, thus, provides us with a way to quantify the impact of decoherence effects on the desired quantum gate. In the numerical simulations we determine the average gate fidelity as a function of evolution time for a given set of parameters characterizing the system. The evolution time for which we find the maximal average fidelity for the desired gate is then chosen as the gate time. Moreover we perform parameters scans within the range of parameters satisfying all the assumptions and requirements discussed in the main text. These scans unveil that choosing the system parameters
\begin{align*}
	\begin{array}{l c l}
			B_z&=&17.525\,\upmu\mathrm{eV},\\
		\frac{A}{2\pi}&=&25\,\mathrm{MHz}, \\
			\frac{\omega_c}{2\pi}&=&4.23\,\mathrm{GHz},\\
\frac{\omega_d}{2\pi}&=&\frac{\omega_c}{2\pi}-\frac{1}{4}\frac{A}{2\pi}+\frac{\delta\omega_d}{2\pi},\\
			\frac{\delta\omega_d}{2\pi}&=&136\,\mathrm{kHz},\\
		b_x&=&0.0517\,\upmu\mathrm{eV},\\
		\frac{g_c}{2\pi}&=&2.418\,\mathrm{MHz},\\
		t_c&=&9\,\upmu\mathrm{eV},\\
		\frac{\epsilon_d}{2\pi}&=&8.46\,\mathrm{MHz},
	\end{array}
\end{align*}
enables a $\sqrt{i\mathrm{SWAP}}$ gate with $\bar{F}_{\mathrm{gate}}=0.90$ and gate time $t_{\sqrt{i\mathrm{SWAP}}}=3.44\,\mathrm{ms}$. For the same choice of parameters we find an $i\mathrm{SWAP}$ gate with $\bar{F}_{\mathrm{gate}}=0.80$ and gate time $t_{i\mathrm{SWAP}}=7.97\,\mathrm{ms}$.

\bibliography{bibliography}

\end{document}